\documentclass[%
reprint,
superscriptaddress,
amsmath,amssymb,
aps,
prb,
]{revtex4-2}
\usepackage{float}
\usepackage{color,graphicx}
\usepackage{ulem}
\graphicspath{ {images/} }
\usepackage{hyperref}
\usepackage{filecontents}

\begin{document}

\title{Anomalous resistivity upturn in Co intercalated TaS$_2$}

\author{Moumita Nandi}
\address{Department of Physics, Indian Institute of Technology Palakkad, Kanjikode, Palakkad, Kerala 678623, India}
\address{Department of Condensed Matter Physics and Materials Science, \\Tata Institute of Fundamental Research, Dr. Homi Bhabha Road, Colaba, Mumbai, 400005, India}
\author{Surajit Dutta}
\address{Department of Condensed Matter Physics and Materials Science, \\Tata Institute of Fundamental Research, Dr. Homi Bhabha Road, Colaba, Mumbai, 400005, India}
\author{ A. Thamizhavel}
\address{Department of Condensed Matter Physics and Materials Science, \\Tata Institute of Fundamental Research, Dr. Homi Bhabha Road, Colaba, Mumbai, 400005, India}
\author{S. K. Dhar}
\address{Department of Condensed Matter Physics and Materials Science, \\Tata Institute of Fundamental Research, Dr. Homi Bhabha Road, Colaba, Mumbai, 400005, India}

\date{\today}

\begin{abstract}
Intercalation of magnetic atoms into the van der Waals gaps of layered transition metal dichalcogenides offers an excellent platform to produce exotic physical properties. Here, we report a detailed study of magnetic and electrical transport properties of Co$_{0.28}$TaS$_2$. The temperature dependent resistivity measurements display anomalous upturn below 11 K, which persists in presence of magnetic field even up to 14 T. In the low temperature region, the resistivity upturn exhibits a unique $T^{1/2}$ scaling behavior, which remains unchanged when an external magnetic field is applied. The $T^{1/2}$ dependence of resistivity upturn is the hallmark of non-Fermi liquid state in orbital two-channel Kondo effect(2CK). This anomalous resistivity upturn in Co$_{0.28}$TaS$_2$ can be attributed to the orbital two-channel Kondo mechanism. 
	
\end{abstract}

\maketitle
\section{INTRODUCTION}

Van der Waals (vdW) materials have captured the attention of the scientific community due to their weakly coupled layered structure that facilitates easily accessible low dimensionality\cite{duong2017van}. Weak van der Waals forces make it easy to separate atomically thin layers from layered solids. In last few decades, layered transition metal dichalcogenides (TMDCs), which belong to the large class of van der Waals (vdW) materials, have attracted enormous interest due to their wide variety of novel physical phenomena such as superconductivity, charge density wave, topological phases, etc.\cite{manzeli20172d,han2018van,garoche1976experimental,nagata1992superconductivity,wilson1975charge,soluyanov2015type,wang2016mote,deng2016experimental,wang2019higher,tamai2016fermi,xu2018evidence}. 
Recently, this interest in bulk properties has been accompanied by a surge of curiosity in atomically thin TMDC layers, which offers a potential platform to explore various technological and fundamental aspects. To expand the diversity in the physical properties of TMDCs, one powerful method is to intercalate atoms between layers. Recently, 3\textit{d}-transition metal intercalated TMDCs have become a rich playground for realizing diverse electrical and magnetic properties. The choice of intercalation atom drastically affects the physical properties in layered TMDCs, such as, M$_x$TaS$_2$ (M=V, Cr, Mn, Fe, Co, Ni) exhibits a wide variety of electronic and magnetic properties depending on the intercalant element. For instance, V$_{0.33}$TaS$_2$\cite{lu2020canted}, Cr$_{0.33}$TaS$_2$\cite{kousaka2016long}, and Mn$_{0.33}$TaS$_2$\cite{zhang2018electrical} are ferromagnetic where as Co$_{0.33}$TaS$_2$\cite{parkin1983magnetic} and Ni$_{0.33}$TaS$_2$\cite{parkin19803} are antiferomagnetic. The concentration of the intercalent elements also plays a crucial role for their various physical properties. For example, Fe$_{x}$TaS$_2$ is ferromagnetic for 0.2 $\leqslant$ $x$ $\leqslant$ 0.4\cite{morosan2007sharp,checkelsky2008anomalous,dijkstra1989band,hardy2015very} and antiferromagnetic for x $>$ 0.4\cite{narita1994preparation}. M$_x$TaS$_2$ also shows novel electronic properties when non-magnetic atoms are intercalated in TaS$_2$. For example, Pb$_{0.33}$TaS$_2$ displays anisotropic superconductivity below 2.8 K and also hosts multiple topological Dirac fermions in the electronic band structure\cite{yang2021anisotropic}. In
intercalated TMDCs, Co$_x$TaS$_2$ is an unique system which exhibits exotic electronic properties in addition to its diverse magnetic properties. Antiferromagnet Co$_{0.33}$TaS$_2$ and ferromagnet Co$_{0.22}$TaS$_2$ both display anomalous Hall effect (AHE) but the origin behind the AHE in Co$_{0.33}$TaS$_2$ is related to its noncollinear magnetic order as well as Weyl crossing in its electronic band structure\cite{park2022field} where as the source is solely ferromagnetic order in Co$_{0.22}$TaS$_2$\cite{liu2021magnetic}. The magnetic as well as electronic properties significantly evolve with Co concentration in Co$_x$TaS$_2$ system (0.22 $\leqslant$ $x$ $\leqslant$ 0.33). In contrast to metallic behavior in Co$_{0.33}$TaS$_2$\cite{park2022field} and Co$_{0.26}$TaS$_2$\cite{liu2022electrical}, semiconductor-like behavior has been reported in Co$_{0.27}$TaS$_2$ that was explained by variable-range hopping mechanism\cite{algaidi2023thickness}. A slight change in Co concentration in Co$_{x}$TaS$_2$ systems, drastically affects the electrical transport properties. Therefore, it would be interesting to investigate other intermediate concentrations of Co intercalation in this system.

In this work, we have synthesized single crystals of Co$_{0.28}$TaS$_2$ and studied its magnetic and electrical transport properties. Here we report anomalous temperature dependence resistivity upturn in Co$_{0.28}$TaS$_2$ at the low temperatures. The upturn scales with $T^{1/2}$ and remains unaffected when an external magnetic field is applied. The unique $T^{1/2}$ dependence of resistivity is the signature of the two-channel Kondo(2CK) effect. Since the pioneering work by Nozi$\acute{e}$res and Blandin\cite{nozieres1980kondo}, the search for two-channel Kondo effect(2CK) in real metallic systems has become a highly sought-after topic. The magnetic field independent $T^{1/2}$ resistivity upturn in Co$_{0.28}$TaS$_2$ can be attributed to the orbital two-channel Kondo(2CK) mechanism.    

\section{METHODS}
 \begin{figure*}[t]
	\includegraphics[width=0.9\textwidth]{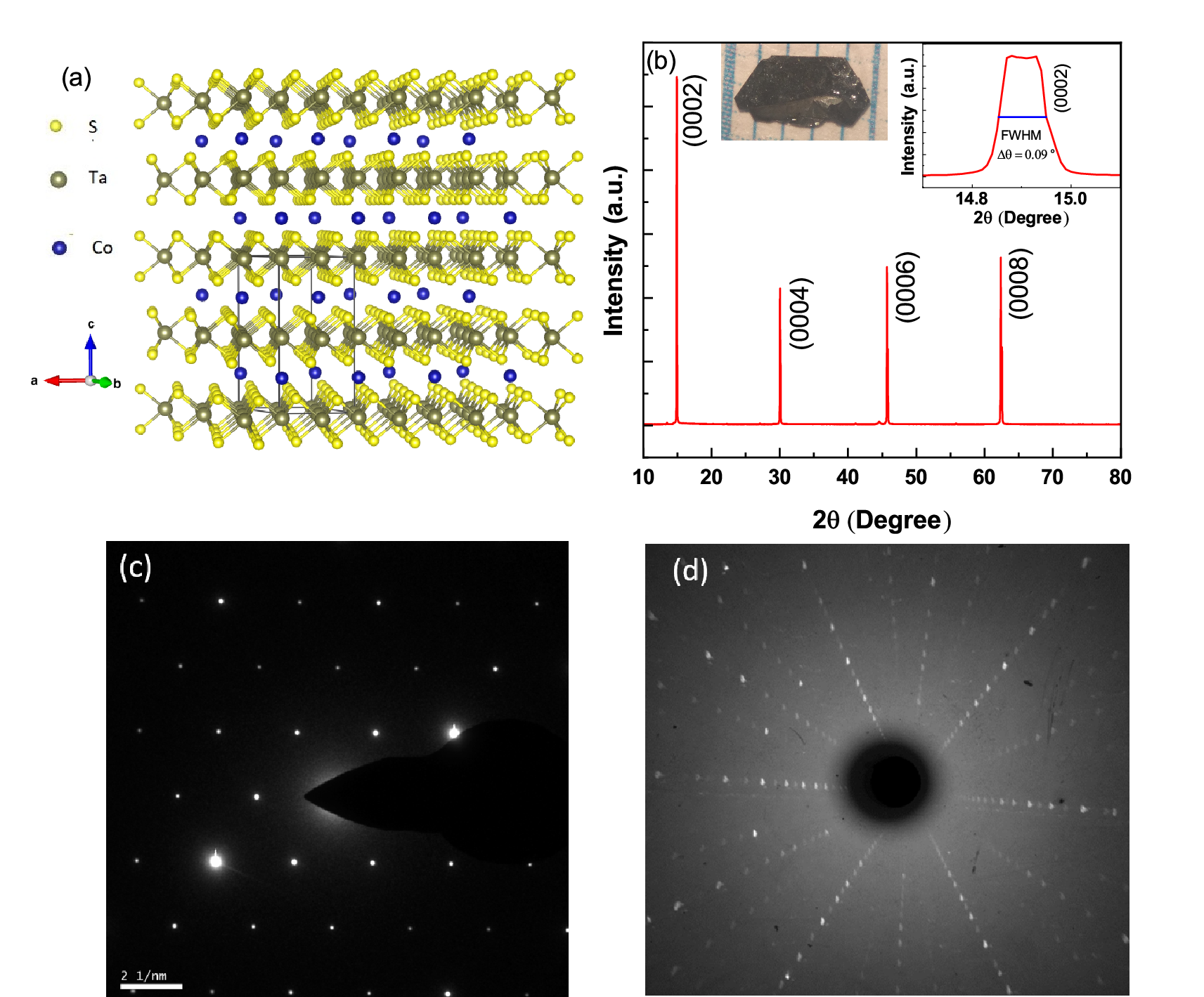}
	\caption{(a) Crystal structure of Co$_x$TaS$_2$. (b) XRD pattern of the single crystal with (000\textit{l}) reflections; the inset of the right panel enlarges the (0002) reflection.(c) SAED diffraction pattern of the Co$_{0.28}$TaS$_2$ crystal along the [001] direction. (d)Laue x-ray pattern of Co$_{0.28}$TaS$_2$ crystal with the sixfold symmetry of the hexagonal structure.}
	\label{Figure1}
\end{figure*}
Single crystals of Co$_{x}$TaS$_2$ were grown by the chemical vapor transport method using iodine as the transport agent. First, a polycrystalline sample was prepared by heating stoichiometric amounts of cobalt powder (Alfa Aesar 99.99\%), tantalum powder (Chemical Center 99. 98\%), and sulfur pieces (Alfa Aesar 99.999\%) in an evacuated silica ampoule at 900 °C for 5 days. Subsequently, 2 g of the powder was loaded together with 0.2 g of iodine in a
silica tube and the tube was evacuated and sealed under vacuum. The ampoule was loaded into a horizontal multizone tube furnace in
which the temperature of the hot zone was kept at 950 °C and the cold zone was kept at 850 °C for 7 days. Shiny plate-like single crystals formed in the cold zone. The typical size of the resulting bulk Co$_{x}$TaS$_2$ single crystals was 4 $\times$ 3 $\times$ 0.2~mm$^3$. The concentration of
Co in synthesized Co$_{x}$TaS$_2$ single crystals was determined to be $x$ = 0.28 by energy-dispersive spectroscopy(EDS). The average Co concentration
was determined by examination of multiple points on multiple samples from the same batch. The EDS data were acquired using a scanning electron microscope (SEM) equipped with an EDS detector. X-ray diffraction measurement on a single crystal sample was done using a a Rigaku diffractometer with a monochromatic Cu-K$_{\rm \alpha}$ radiation ($\lambda$ =0.15418 nm). Selected area electron diffraction (SAED) was performed at room temperature, placing the single crystal on a transmission electron microscope (TEM) grid.  Magnetic measurements were performed using a SQUID magnetometer (Quantum Design). Heat capacity and electrical measurements were performed using a physical property measurement system (PPMS).

\section{RESULTS AND DISCUSSION}
\begin{figure*}[t]
	\includegraphics[width=0.9\textwidth]{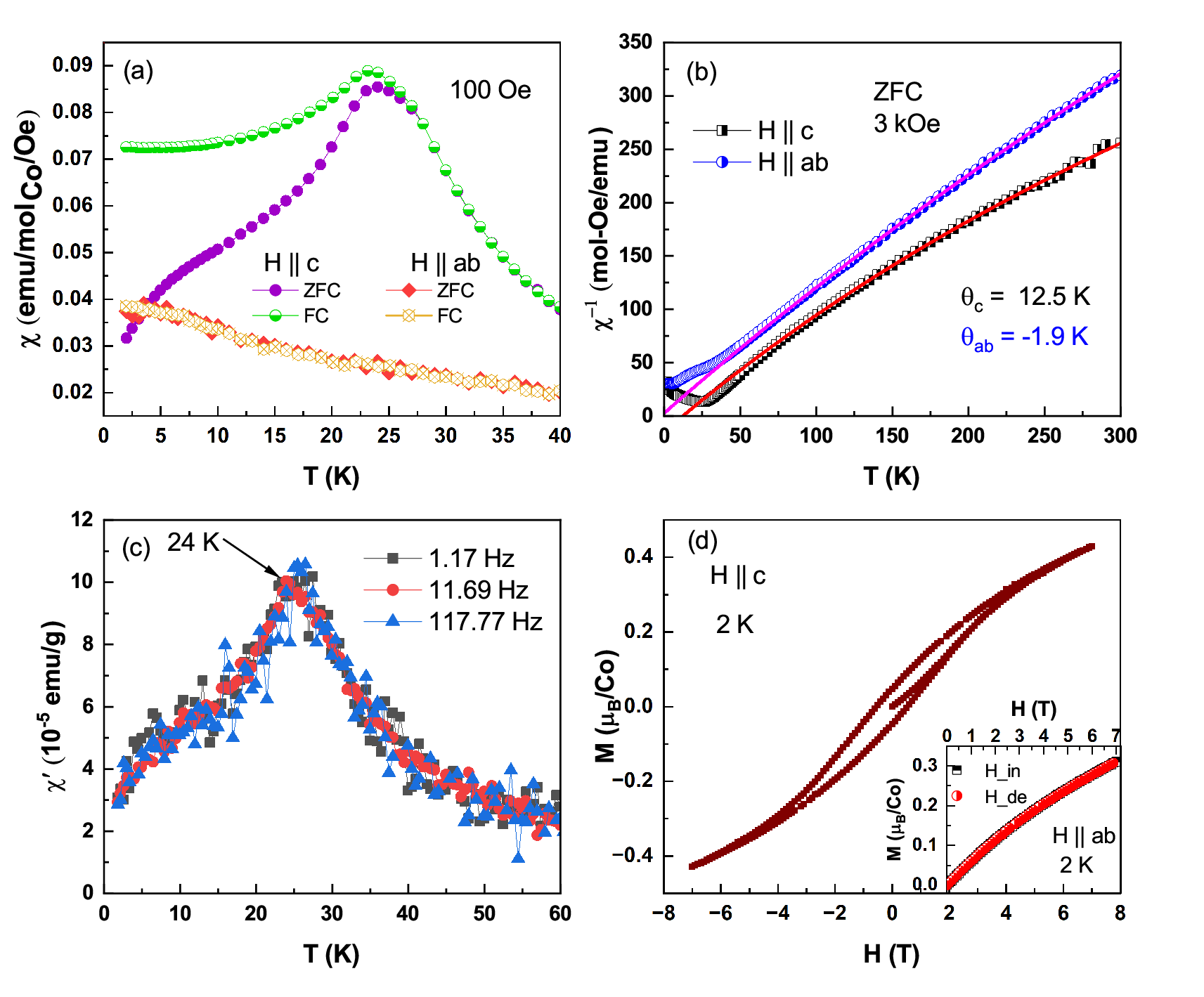}
	\caption{(a) Temperature dependence of magnetic susceptibility $\chi (T)$ in both ZFC and FC modes measured at 100 Oe for magnetic field $H \parallel c$ and $H \parallel ab$. (b) Inverse susceptibility 1/$\chi (T)$ at 3 kOe for magnetic field $H \parallel c$ and $H \parallel ab$ fitted by the Curie-Weiss law (solid lines). (c) Temperature dependence of the ac magnetic susceptibility. (d) Field-dependent magnetization with both the $H \parallel c$ and $H \parallel ab$ at 2 K.}
	\label{Figure2}
\end{figure*}
Figure~\ref{Figure1}(a) presents the hexagonal crystal structure of Co$_{x}$TaS$_2$, which crystallizes in the space group \textit{P}6$_3$22.  TaS$_2$ is a quasi-two-dimensional system with van der Waals interaction between the layers. Magnetic Co atoms are intercalated between TaS$_2$ layers.
Figure~\ref{Figure1}(b) shows the XRD pattern of Co$_{0.28}$TaS$_2$
single crystal. The sharp peaks in the XRD $\theta$-2$\theta$ scan can be indexed with (000l) planes, indicating that the crystallographic $c$ axis is perpendicular to the surface of the plate, as shown in Figure~\ref{Figure1}(b). The inset of Figure~\ref{Figure1}(b) shows that the full width at half-maximum (FWHM) of the (0002) peak is only 0.09$^{\circ}$ which indicates the high crystalline quality. The value of the lattice parameter $c$ has been extracted from this XRD data by using Bragg's law. The extracted lattice parameter ($c$ = 11.89 \AA) is smaller campared with Co$_{0.33}$TaS$_2$ ($c$ = 11.93 \AA)\cite{parkin1983magnetic,van1971magnetic}, which is consistent with lesser amount of Co intercalation in Co$_{0.28}$TaS$_2$. Figure~\ref{Figure1} (c) shows the hexagonal arranged diffraction spots in the selected area electron diffraction (SAED) pattern of Co$_{0.28}$TaS$_2$ along the [001] direction. The Laue diffraction pattern of a Co$_{0.28}$TaS$_2$ crystal also confirms the six-fold symmetry of the hexagonal structure of the (0001)-plane, as shown in Figure~\ref{Figure1}(d).

\begin{figure*}
	\includegraphics[width=0.8\textwidth]{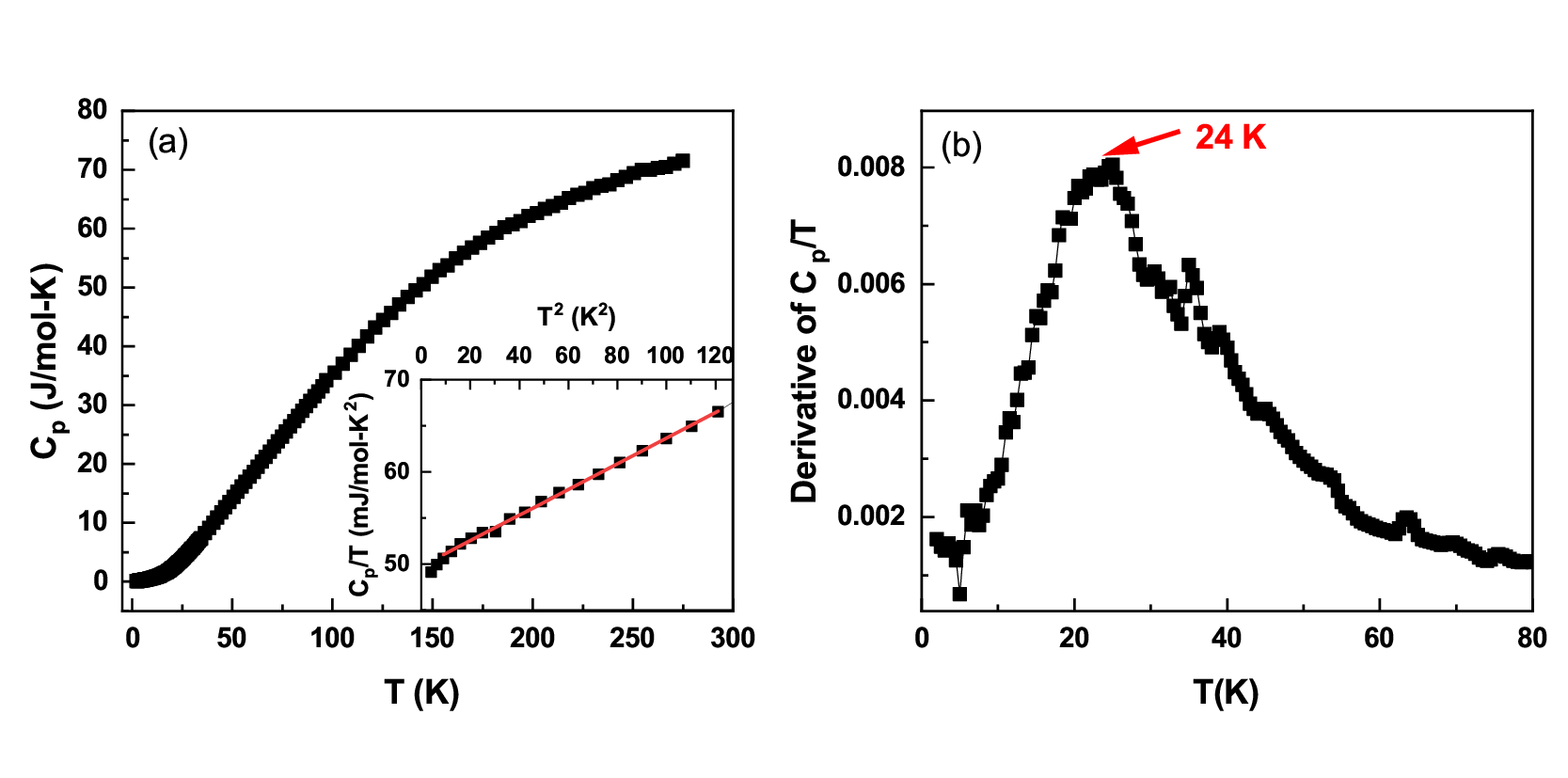}
	\caption{(a) Heat capacity C$_p$ versus Temperature ($T$) plot; inset shows low-temperature C$_p$/$T$ vs $T$$^2$ data fitted by $C_p$/$T$ = $\gamma$ + $\beta$ $T^2$. (b) Derivative of C$_p$/$T$ vs $T$.}
	\label{Figure3}
\end{figure*}
 
To gain insight into the magnetic properties of Co$_{0.28}$TaS$_2$, we conducted temperature-dependent magnetization measurements for magnetic field $H$ parallel to the $c$-axis and the $ab$ plane respectively, using both zero-field-cooled (ZFC) and field-cooled (FC) protocols. Figure~\ref{Figure2}(a) displays the temperature dependence of magnetic susceptibility $\chi (T)$ measured at $H$ = 100 Oe. The susceptibility $\chi (T)$ for $H \parallel c$ is higher than that for $H \parallel ab$, suggesting that the magnetic easy axis is parallel to the $c$ axis. $\chi (T)$ for $H \parallel c$ shows a peak around 24 K, while $\chi (T)$ for $H \parallel ab$ increases down to 2 K as the temperature decreases. Additionally, a bifurcation in the ZFC and FC curves has been observed only for $H \parallel c$, which is a phenomenon also seen in anisotropic spin-glass materials\cite{dragomir2019comparing}. To get more information about the bifurcation, we have measured ac magnetic susceptibility at different frequencies. The ac susceptibility versus temperature measurements [Figure~\ref{Figure2}(c)] do not show any frequency dependence, eleminating the possibility of spin-glass. We have also plotted the temperature dependence of inverse susceptibility 1/$\chi (T)$ measured in an applied magnetic field of $3$~kOe, as shown in Figure~\ref{Figure2}(b), along the two principal crystallographic directions, namely $H \parallel c$ and $H \parallel ab$. In the temperature range 100-300 K, 1/$\chi (T)$ data can be well fitted by the modified Curie-Weiss law, $\chi$ = $\chi$$_0$ + $C$/($T$$-$ $\theta$), where $\chi$$_0$ is a temperature independent term, and $C$ and $\theta$ are the Curie-Weiss constant and Weiss temperature, respectively. The Weiss temperature is $\theta$$_c$ = 12.5 K for $H \parallel c$ and $\theta$$_{ab}$ = $-$1.9 K for $H \parallel ab$,
indicating the dominance of ferromagnetic (FM) exchange interactions along the $c$ axis and antiferromagnetic (AFM) in the $ab$ plane. The derived effective moment is 2.59 $\mu_B$/Co for $H \parallel c$ and 2.53 $\mu_B$/Co for $H \parallel ab$, which is smaller than the spin-only moment of 3.87 $\mu_B$ for Co$^{2+}$. The loss of Co magnetic moment is in line with previous reports of  other Co$_{x}$TaS$_2$ systems\cite{liu2021magnetic,liu2022electrical}. The mixing interaction of 3\textit{d} Co electrons and Ta conduction electrons can be possible cause of this loss. Similar result has also been found in Co$_{0.33}$NbS$_2$ by neutron scattering study\cite{parkin1983magnetic}. Figure~\ref{Figure2}(d) presents the field-dependent magnetization measured at 2 K. The hysteresis loop has clearly been observed when magnetic field is applied along the $c$ axis. The $M$($H$) curves for the $H \parallel ab$ do not show any hysteresis[Inset of Figure~\ref{Figure2}(d)]. At $2$~K, well below the Curie temperature, $M$($H$) curve is non-linear for $H \parallel ab$  and $M$($H$) curve for $H \parallel c$ is far away from perfect square-like hysteresis loop which may arise from noncollinear arrangement of magnetic moments. Previous studies have already reported noncolliniear antiferromagnetic structure in Co$_{0.33}$TaS$_2$\cite{parkin1983magnetic,park2022field}. Very recently, neutron diffraction measurements on Co$_{x}$TaS$_2$ reveal that magnetic behavior changes drastically with slight variation of Co concentration\cite{park2024composition}. Different magnetic ground states emerge depending on the Co concentration. For example, the magnetic ground state
with wave vector q =(1/3, 0, 0) has been reported for 0.330 $<$ $x$ $<$ 0.34, where a weak ferromagnetic moment emerges with magnetic ordering characterized by the wave vector q =(1/2, 0, 0) for 0.299 $<$ $x$ $<$ 0.325. A recent report of a tetrahedral triple-Q ordering scenario explains quite well the coexsistance of long-range antiferromagnetic ordering with a weak ferromagnetic moment\cite{park2023tetrahedral}. In Co$_{0.28}$TaS$_2$, though we observe an antiferromagnetic-like transition around 24 K for $H \parallel c$, the bifurcation of FC-ZFC and the hysteresis in the M-H loop indicate the signature of a small ferromagnetic component. Such kind of magnetic behavior may arise from a canted antiferromagnetic state. Detailed microscopic investigations are required to obtain the exact picture of the magnetic ground state of Co$_{0.28}$TaS$_2$.  

To understand the transition found around 24 K in more detail, we have performed a temperature-dependent heat capacity ($C_p$) measurement. Figure~\ref{Figure3}(a) shows no peak in the $C_p(T)$ versus temperature ($T$) plot down to 2 K. However, a very broad peak has been observed around 24 K in derivative of the $C_p(T)$/$T$ versus $T$ plot, shown in Figure~\ref{Figure3}(b), indicating the signature of the transition found at 24 K in the magnetization data in heat capacity measurement. The exchange coupling between intercalated-Co moments in the TaS$_2$ layer may arise from the Ruderman-
Kittel-Kasuya-Yosida (RKKY) interaction\cite{aristov1997indirect,yosida1957magnetic}. The low temperature data from 2.5 to 10 K can be well fitted by $C_p$/$T$ = $\gamma$ + $\beta$ $T^2$, shown in the inset of Figure~\ref{Figure3}(a), where the first term is the Sommerfeld electronic specific heat coefficient and the second term is the low temperature limit of the lattice heat capacity. The derived $\gamma$ is 49.7 mJ mol$^{-1}$ K$^{-2}$ which is very close to the value previously reported for Co$_{0.26}$TaS$_2$\cite{liu2022electrical}. 
\begin{figure*}[t]
	\includegraphics[width=1\textwidth]{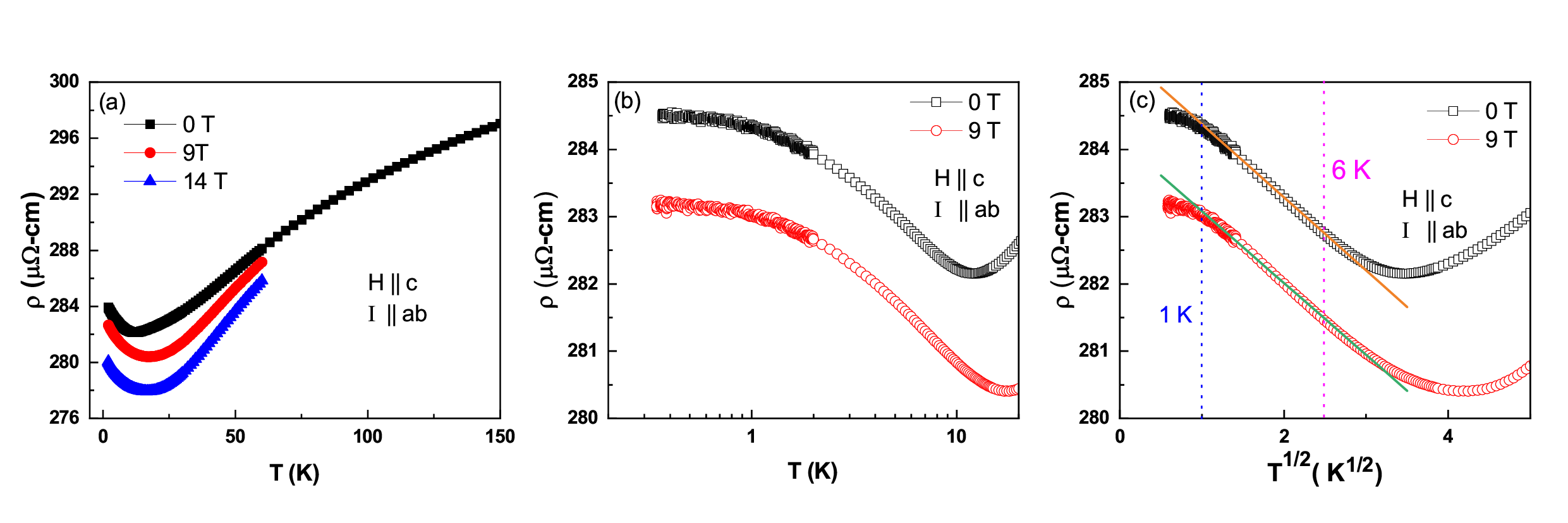}
	\caption{Low-temperature electrical resistivity upturn.(a) Temperature dependent resistivity ($\rho$ ) measurement at various magnetic fields.(b) Semilog plot of $\rho$ versus $T$,(c) $\rho$ versus $T^{1/2}$}
	\label{Figure4}
\end{figure*}

 \begin{figure*}[t]
 	\includegraphics[width=0.8\textwidth]{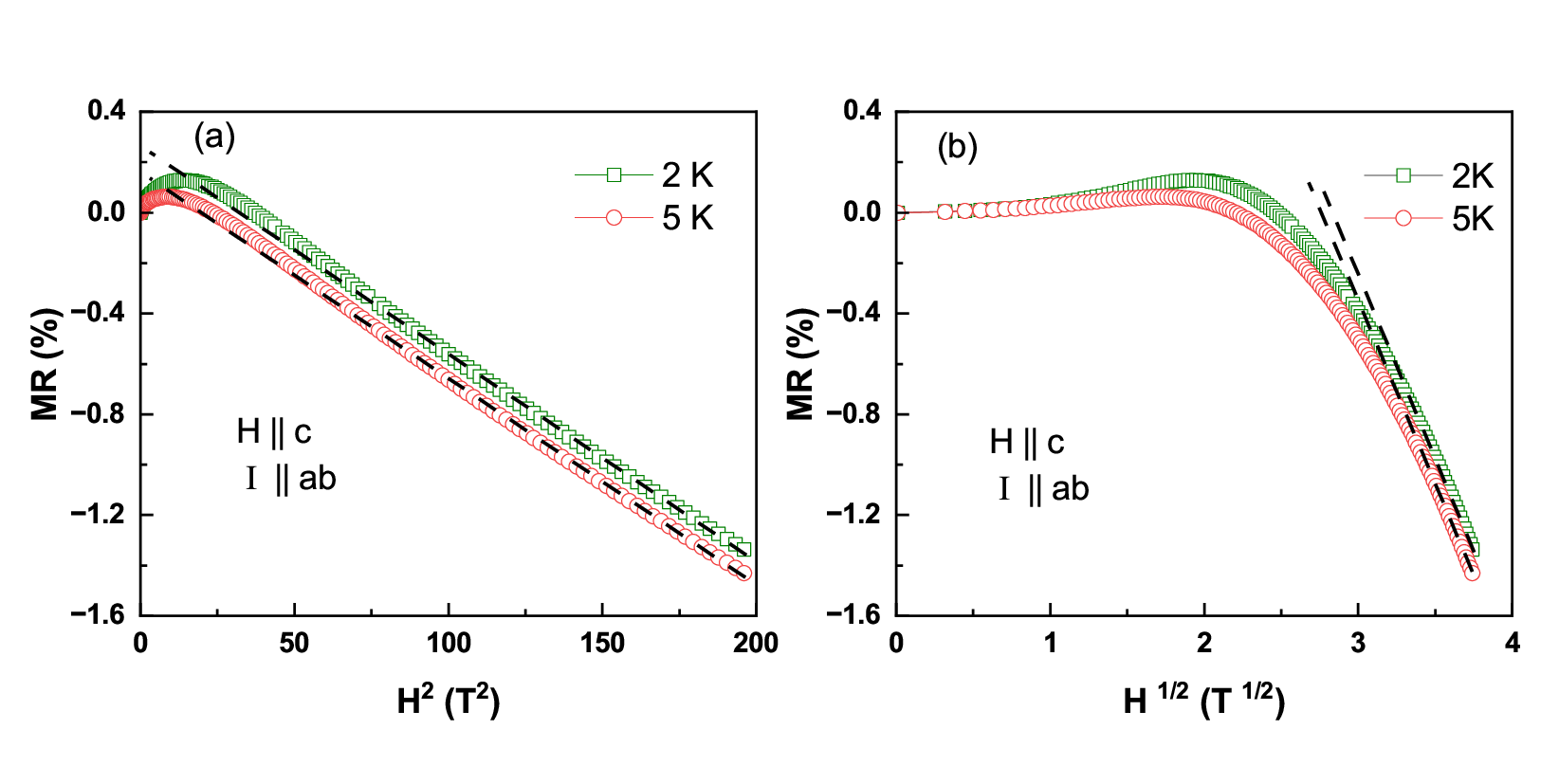}
 	\caption{Magnetoresistance (MR) plots at different temperatures.(a) MR versus $H^2$,(b) MR versus $H^{1/2}$}
 	\label{Figure5}
 \end{figure*}
We have also performed temperature-dependent resistivity measurements to investigate the electrical transport properties of Co$_{0.28}$TaS$_2$. Figure~\ref{Figure4}(a) shows the temperature dependence of the longitudinal resistivity ($\rho$) both in the presence and absence of external magnetic field. The plot $\rho$ vs. $T$ exhibits a minimum of around 11 K at $H$ = 0 T. Below this minimum, $\rho$ shows an upturn. Similar upturn behavior has been observed in the presence of a magnetic field at least up to 14 T when the magnetic field is applied along the $c$ axis and the current (I) is parallel to the $ab$ plane. Sometimes, an upturn in low temperature resistivity is observed due to the single-channel Kondo (1CK) effect when conduction electrons are scattered by dilute magnetic impurities present in a metal\cite{hewson1997kondo,kondo1964resistance,kondo1970theory}. To comprehend the source of the resistivity upturn, we have measured the resistivity $\rho (T)$ down to the milli Kelvin range, as shown in Figure~\ref{Figure4} (b) and (c). In the case of single channel Kondo (1CK), $\rho$ displays a logarithmic dependence with temperature\cite{nagai1975resistivity}. However, Figure~\ref{Figure4}(b) and (c) reveal that $\rho$ significantly deviates from the $\ln T$ behavior where as it exhibits $T^{1/2}$ dependence. Additionally, the low temperature resistivity upturn remains unaffected in presence of the magnetic field. Those observations suggest that single channel Kondo mechanism can not be the cause of resistivity upturn. Below 1 K, the resistivity upturn deviates from the non-Fermi liquid $T^{1/2}$ behavior, indicating the crossover to Fermi liquid behavior upon further cooling.

$H$-independent $T^{1/2}$ resistivity upturn can have two possible origins (i) Electron electron interaction (EEI) in three-dimensional disordered metal\cite{altshuler1985electron, lee1985disordered} (ii) Two-channel Kondo effect (2CK). In the case of EEI, magnetoregistance should exhibit $H^{1/2}$ dependence in the high magnetic field region, which is a characteristic signature for disordered systems\cite{lee1985disordered}. Figure~\ref{Figure5} displays the magnetoresistance (MR) vs. $H$ plot at low temperature. Figure~\ref{Figure5} (b) clearly shows that there is no $H^{1/2}$ dependence while the MR curves demonstrate $H^2$ dependence in high magnetic field region [Figure~\ref{Figure5}(a)]. Dashed lines will guide one to observe the field dependence of magnetoresistance. Therefore, the electron-electron interaction is probably not responsible for resistivity upturn in Co$_{0.28}$TaS$_2$. 

Another possible mechanism for a resistivity upturn is the two-channel Kondo effect. The two-channel Kondo effect can occur when there are two degenerate channels of conduction electrons that interact with an impurity spin\cite{nozieres1980kondo}. Very few bulk systems were reported to exhibit $H$-independent $T^{1/2}$ resistivity upturn due to 2CK. The magnetic field independent $T^{1/2}$ resistivity upturn in ThAsSe\cite{cichorek2005two}, ZrAs$_{1.58}$Se$_{0.39}$\cite{cichorek2016two}, can be well explained by the orbital two-channel Kondo effect. Apart from non-magnetic systems, resistivity upturn due to orbital two-channel Kondo effect (2CK) has also been observed in the ferromagnetic films L1$_0$-MnAl\cite{zhu2016orbital} and L1$_0$-MnGa\cite{zhu2016observation}. The orbital 2CK is proposed to arise in structural two-level systems (TLS) when pseudospin-1/2 is equally coupled to two spin channels of conduction electrons through resonant scattering centers that possess orbital degrees of freedom\cite{zawadowski1980kondo,cox1997exotic,von19982}.  Co-intercalation in TaS$_2$ may generate structural two-level system (TLS). The signature of TLS in Co$_{0.28}$TaS$_2$ has been observed in low-temperature specific heat results.The extracted value of the Sommerfeld coefficient ($\gamma$ = 49.7 mJ mol$^{-1}$ K$^{-2}$) is 5 times larger compared to TaS$_2$ ($\gamma$ = 8.8 mJ mol$^{-1}$ K$^{-2}$)\cite{abdel2016enhancement}. Such a large value of the Sommerfeld coefficient is comparable to that of heavy-fermion systems. Earlier theoretical studies reported a linear temperature dependence of the specific heat in two-level systems (TLS) at low temperature\cite{anderson1972anomalous,phillips1972tunneling}. Most likely, an additional linear-in-$T$ term adds to the specific heat due to the formation of TLS in Co$_{0.28}$TaS$_2$.  

Previous studies suggest that M intercalators tend to form ordered superstructures of 2 $\times$ 2 or $\sqrt{3}$ $\times$ $\sqrt{3}$ ordered superstructures in M$_{x}$TaS$_2$ systems\cite{parkin19803}. For example, a 2 $\times$ 2 superstructure near $x$ = 1/4\cite{morosan2007sharp} and a $\sqrt{3}$ $\times$ $\sqrt{3}$ superstructure near $x$ = 1/3\cite{morosan2007sharp,hardy2015very,chen2016correlations} are reported in Fe$_{x}$TaS$_2$. Furthermore, the $\sqrt{3}$ $\times$ $\sqrt{3}$ superstructure with vacancies is reported for in between intercalation of  $x$=1/4 and $x$=1/3\cite{hardy2015very}. In Co$_{0.33}$TaS$_2$, formation of $\sqrt{3}$ $\times$ $\sqrt{3}$ supercell with homogeneous Co intercalation has been realized by single crystal XRD\cite{park2022field}. Therefore, just like Fe$_{x}$TaS$_2$, the intermediate Co concentration ($x$ $\sim$ 0.28) may create vacancies in the superstructure arrangement of Co$_{x}$TaS$_2$ system also. The vacancies at the As site in ZrAs$_{1.58}$Se$_{0.39}$\cite{cichorek2016two} and HfAs$_{1.7}$Se$_{0.2}$\cite{czulucki2010crystal} have been reported to be a possible cause behind the formation of the structural two-level system(TLS). Theoretical studies have revealed that the tunneling of defects with two energy levels can lead to the Kondo effect in TLS\cite{zawadowski1980kondo,vladar1983theory1,vladar1983theory2}. Particularly, the $H$-independent $T^{1/2}$ resistivity upturn in ZrAs$_{1.58}$Se$_{0.39}$, has been well explained by a model where the vacancies in the square nets of As layers can create dynamic structural scattering centers which ultimately gives arise to an orbital two-channel Kondo effect (2CK) in this compound\cite{cichorek2016two,cichorek2017cichorek}. Here, we propose that the formation of Co vacancies in the ordered superstructure is a probable scenario, which in turn can produce the orbital two-channel Kondo effect (2CK) in Co$_{0.28}$TaS$_2$. Therefore, the orbital two-channel Kondo effect (2CK) can be a possible origin of the anomalous resistivity upturn in Co$_{0.28}$TaS$_2$. 

\section{CONCLUSIONS}
In summary, we synthesized single crystals of Co$_{0.28}$TaS$_2$ by intercalating $x$ $\sim$ 0.28 Co into the van der Waals gaps of TaS$_2$. We have systematically investigated the magnetic and transport properties of Co$_{0.28}$TaS$_2$. Highly anisotropic magnetic properties have been observed in magnetization studies. In electrical transport properties, we first report an anomolous upturn in temperature-dependent resistivity measurements that exhibit $T^{1/2}$ dependence at low temperature. This unique $T^{1/2}$ behavior remains unaffected in the presence of an external magnetic field. The magnetic field independent resistivity upturn is probably originated because of the orbital two channel Kondo (2CK) effect. The orbital 2CK in Co$_{0.28}$TaS$_2$ can be explained by a possible scenario where a structural two-level system is formed due to vacancies in the ordered superstructure of Co intercalators. 

\section{ACKNOWLEDGMENTS}
We acknowledge Department of Science and Technology (DST) India (Innovation in Science Pursuit for Inspired Research-INSPIRE Faculty Grant). We thank Prof. Pratap Raychaudhuri for his help performing experiments in the milli-Kelvin temperature range.

 %\bibliography{mybibfile}
\bibliography{Ref}  

\begin{thebibliography}{55}%
\makeatletter
\providecommand \@ifxundefined [1]{%
 \@ifx{#1\undefined}
}%
\providecommand \@ifnum [1]{%
 \ifnum #1\expandafter \@firstoftwo
 \else \expandafter \@secondoftwo
 \fi
}%
\providecommand \@ifx [1]{%
 \ifx #1\expandafter \@firstoftwo
 \else \expandafter \@secondoftwo
 \fi
}%
\providecommand \natexlab [1]{#1}%
\providecommand \enquote  [1]{``#1''}%
\providecommand \bibnamefont  [1]{#1}%
\providecommand \bibfnamefont [1]{#1}%
\providecommand \citenamefont [1]{#1}%
\providecommand \href@noop [0]{\@secondoftwo}%
\providecommand \href [0]{\begingroup \@sanitize@url \@href}%
\providecommand \@href[1]{\@@startlink{#1}\@@href}%
\providecommand \@@href[1]{\endgroup#1\@@endlink}%
\providecommand \@sanitize@url [0]{\catcode `\\12\catcode `\$12\catcode
  `\&12\catcode `\#12\catcode `\^12\catcode `\_12\catcode `\%12\relax}%
\providecommand \@@startlink[1]{}%
\providecommand \@@endlink[0]{}%
\providecommand \url  [0]{\begingroup\@sanitize@url \@url }%
\providecommand \@url [1]{\endgroup\@href {#1}{\urlprefix }}%
\providecommand \urlprefix  [0]{URL }%
\providecommand \Eprint [0]{\href }%
\providecommand \doibase [0]{https://doi.org/}%
\providecommand \selectlanguage [0]{\@gobble}%
\providecommand \bibinfo  [0]{\@secondoftwo}%
\providecommand \bibfield  [0]{\@secondoftwo}%
\providecommand \translation [1]{[#1]}%
\providecommand \BibitemOpen [0]{}%
\providecommand \bibitemStop [0]{}%
\providecommand \bibitemNoStop [0]{.\EOS\space}%
\providecommand \EOS [0]{\spacefactor3000\relax}%
\providecommand \BibitemShut  [1]{\csname bibitem#1\endcsname}%
\let\auto@bib@innerbib\@empty
%</preamble>
\bibitem [{\citenamefont {Duong}\ \emph {et~al.}(2017)\citenamefont {Duong},
  \citenamefont {Yun},\ and\ \citenamefont {Lee}}]{duong2017van}%
  \BibitemOpen
  \bibfield  {author} {\bibinfo {author} {\bibfnamefont {D.~L.}\ \bibnamefont
  {Duong}}, \bibinfo {author} {\bibfnamefont {S.~J.}\ \bibnamefont {Yun}},\
  and\ \bibinfo {author} {\bibfnamefont {Y.~H.}\ \bibnamefont {Lee}},\
  }\href@noop {} {\bibfield  {journal} {\bibinfo  {journal} {ACS nano}\
  }\textbf {\bibinfo {volume} {11}},\ \bibinfo {pages} {11803} (\bibinfo {year}
  {2017})}\BibitemShut {NoStop}%
\bibitem [{\citenamefont {Manzeli}\ \emph {et~al.}(2017)\citenamefont
  {Manzeli}, \citenamefont {Ovchinnikov}, \citenamefont {Pasquier},
  \citenamefont {Yazyev},\ and\ \citenamefont {Kis}}]{manzeli20172d}%
  \BibitemOpen
  \bibfield  {author} {\bibinfo {author} {\bibfnamefont {S.}~\bibnamefont
  {Manzeli}}, \bibinfo {author} {\bibfnamefont {D.}~\bibnamefont
  {Ovchinnikov}}, \bibinfo {author} {\bibfnamefont {D.}~\bibnamefont
  {Pasquier}}, \bibinfo {author} {\bibfnamefont {O.~V.}\ \bibnamefont
  {Yazyev}},\ and\ \bibinfo {author} {\bibfnamefont {A.}~\bibnamefont {Kis}},\
  }\href@noop {} {\bibfield  {journal} {\bibinfo  {journal} {Nature Reviews
  Materials}\ }\textbf {\bibinfo {volume} {2}},\ \bibinfo {pages} {1} (\bibinfo
  {year} {2017})}\BibitemShut {NoStop}%
\bibitem [{\citenamefont {Han}\ \emph {et~al.}(2018)\citenamefont {Han},
  \citenamefont {Duong}, \citenamefont {Keum}, \citenamefont {Yun},\ and\
  \citenamefont {Lee}}]{han2018van}%
  \BibitemOpen
  \bibfield  {author} {\bibinfo {author} {\bibfnamefont {G.~H.}\ \bibnamefont
  {Han}}, \bibinfo {author} {\bibfnamefont {D.~L.}\ \bibnamefont {Duong}},
  \bibinfo {author} {\bibfnamefont {D.~H.}\ \bibnamefont {Keum}}, \bibinfo
  {author} {\bibfnamefont {S.~J.}\ \bibnamefont {Yun}},\ and\ \bibinfo {author}
  {\bibfnamefont {Y.~H.}\ \bibnamefont {Lee}},\ }\href@noop {} {\bibfield
  {journal} {\bibinfo  {journal} {Chemical reviews}\ }\textbf {\bibinfo
  {volume} {118}},\ \bibinfo {pages} {6297} (\bibinfo {year}
  {2018})}\BibitemShut {NoStop}%
\bibitem [{\citenamefont {Garoche}\ \emph {et~al.}(1976)\citenamefont
  {Garoche}, \citenamefont {Veyssie}, \citenamefont {Manuel},\ and\
  \citenamefont {Molini{\'e}}}]{garoche1976experimental}%
  \BibitemOpen
  \bibfield  {author} {\bibinfo {author} {\bibfnamefont {P.}~\bibnamefont
  {Garoche}}, \bibinfo {author} {\bibfnamefont {J.}~\bibnamefont {Veyssie}},
  \bibinfo {author} {\bibfnamefont {P.}~\bibnamefont {Manuel}},\ and\ \bibinfo
  {author} {\bibfnamefont {P.}~\bibnamefont {Molini{\'e}}},\ }\href@noop {}
  {\bibfield  {journal} {\bibinfo  {journal} {Solid State Communications}\
  }\textbf {\bibinfo {volume} {19}},\ \bibinfo {pages} {455} (\bibinfo {year}
  {1976})}\BibitemShut {NoStop}%
\bibitem [{\citenamefont {Nagata}\ \emph {et~al.}(1992)\citenamefont {Nagata},
  \citenamefont {Aochi}, \citenamefont {Abe}, \citenamefont {Ebisu},
  \citenamefont {Hagino}, \citenamefont {Seki},\ and\ \citenamefont
  {Tsutsumi}}]{nagata1992superconductivity}%
  \BibitemOpen
  \bibfield  {author} {\bibinfo {author} {\bibfnamefont {S.}~\bibnamefont
  {Nagata}}, \bibinfo {author} {\bibfnamefont {T.}~\bibnamefont {Aochi}},
  \bibinfo {author} {\bibfnamefont {T.}~\bibnamefont {Abe}}, \bibinfo {author}
  {\bibfnamefont {S.}~\bibnamefont {Ebisu}}, \bibinfo {author} {\bibfnamefont
  {T.}~\bibnamefont {Hagino}}, \bibinfo {author} {\bibfnamefont
  {Y.}~\bibnamefont {Seki}},\ and\ \bibinfo {author} {\bibfnamefont
  {K.}~\bibnamefont {Tsutsumi}},\ }\href@noop {} {\bibfield  {journal}
  {\bibinfo  {journal} {Journal of Physics and Chemistry of Solids}\ }\textbf
  {\bibinfo {volume} {53}},\ \bibinfo {pages} {1259} (\bibinfo {year}
  {1992})}\BibitemShut {NoStop}%
\bibitem [{\citenamefont {Wilson}\ \emph {et~al.}(1975)\citenamefont {Wilson},
  \citenamefont {Di~Salvo},\ and\ \citenamefont {Mahajan}}]{wilson1975charge}%
  \BibitemOpen
  \bibfield  {author} {\bibinfo {author} {\bibfnamefont {J.~A.}\ \bibnamefont
  {Wilson}}, \bibinfo {author} {\bibfnamefont {F.}~\bibnamefont {Di~Salvo}},\
  and\ \bibinfo {author} {\bibfnamefont {S.}~\bibnamefont {Mahajan}},\
  }\href@noop {} {\bibfield  {journal} {\bibinfo  {journal} {Advances in
  Physics}\ }\textbf {\bibinfo {volume} {24}},\ \bibinfo {pages} {117}
  (\bibinfo {year} {1975})}\BibitemShut {NoStop}%
\bibitem [{\citenamefont {Soluyanov}\ \emph {et~al.}(2015)\citenamefont
  {Soluyanov}, \citenamefont {Gresch}, \citenamefont {Wang}, \citenamefont
  {Wu}, \citenamefont {Troyer}, \citenamefont {Dai},\ and\ \citenamefont
  {Bernevig}}]{soluyanov2015type}%
  \BibitemOpen
  \bibfield  {author} {\bibinfo {author} {\bibfnamefont {A.~A.}\ \bibnamefont
  {Soluyanov}}, \bibinfo {author} {\bibfnamefont {D.}~\bibnamefont {Gresch}},
  \bibinfo {author} {\bibfnamefont {Z.}~\bibnamefont {Wang}}, \bibinfo {author}
  {\bibfnamefont {Q.}~\bibnamefont {Wu}}, \bibinfo {author} {\bibfnamefont
  {M.}~\bibnamefont {Troyer}}, \bibinfo {author} {\bibfnamefont
  {X.}~\bibnamefont {Dai}},\ and\ \bibinfo {author} {\bibfnamefont {B.~A.}\
  \bibnamefont {Bernevig}},\ }\href@noop {} {\bibfield  {journal} {\bibinfo
  {journal} {Nature}\ }\textbf {\bibinfo {volume} {527}},\ \bibinfo {pages}
  {495} (\bibinfo {year} {2015})}\BibitemShut {NoStop}%
\bibitem [{\citenamefont {Wang}\ \emph {et~al.}(2016)\citenamefont {Wang},
  \citenamefont {Gresch}, \citenamefont {Soluyanov}, \citenamefont {Xie},
  \citenamefont {Kushwaha}, \citenamefont {Dai}, \citenamefont {Troyer},
  \citenamefont {Cava},\ and\ \citenamefont {Bernevig}}]{wang2016mote}%
  \BibitemOpen
  \bibfield  {author} {\bibinfo {author} {\bibfnamefont {Z.}~\bibnamefont
  {Wang}}, \bibinfo {author} {\bibfnamefont {D.}~\bibnamefont {Gresch}},
  \bibinfo {author} {\bibfnamefont {A.~A.}\ \bibnamefont {Soluyanov}}, \bibinfo
  {author} {\bibfnamefont {W.}~\bibnamefont {Xie}}, \bibinfo {author}
  {\bibfnamefont {S.}~\bibnamefont {Kushwaha}}, \bibinfo {author}
  {\bibfnamefont {X.}~\bibnamefont {Dai}}, \bibinfo {author} {\bibfnamefont
  {M.}~\bibnamefont {Troyer}}, \bibinfo {author} {\bibfnamefont {R.~J.}\
  \bibnamefont {Cava}},\ and\ \bibinfo {author} {\bibfnamefont {B.~A.}\
  \bibnamefont {Bernevig}},\ }\href@noop {} {\bibfield  {journal} {\bibinfo
  {journal} {Physical review letters}\ }\textbf {\bibinfo {volume} {117}},\
  \bibinfo {pages} {056805} (\bibinfo {year} {2016})}\BibitemShut {NoStop}%
\bibitem [{\citenamefont {Deng}\ \emph {et~al.}(2016)\citenamefont {Deng},
  \citenamefont {Wan}, \citenamefont {Deng}, \citenamefont {Zhang},
  \citenamefont {Ding}, \citenamefont {Wang}, \citenamefont {Yan},
  \citenamefont {Huang}, \citenamefont {Zhang}, \citenamefont {Xu} \emph
  {et~al.}}]{deng2016experimental}%
  \BibitemOpen
  \bibfield  {author} {\bibinfo {author} {\bibfnamefont {K.}~\bibnamefont
  {Deng}}, \bibinfo {author} {\bibfnamefont {G.}~\bibnamefont {Wan}}, \bibinfo
  {author} {\bibfnamefont {P.}~\bibnamefont {Deng}}, \bibinfo {author}
  {\bibfnamefont {K.}~\bibnamefont {Zhang}}, \bibinfo {author} {\bibfnamefont
  {S.}~\bibnamefont {Ding}}, \bibinfo {author} {\bibfnamefont {E.}~\bibnamefont
  {Wang}}, \bibinfo {author} {\bibfnamefont {M.}~\bibnamefont {Yan}}, \bibinfo
  {author} {\bibfnamefont {H.}~\bibnamefont {Huang}}, \bibinfo {author}
  {\bibfnamefont {H.}~\bibnamefont {Zhang}}, \bibinfo {author} {\bibfnamefont
  {Z.}~\bibnamefont {Xu}}, \emph {et~al.},\ }\href@noop {} {\bibfield
  {journal} {\bibinfo  {journal} {Nature Physics}\ }\textbf {\bibinfo {volume}
  {12}},\ \bibinfo {pages} {1105} (\bibinfo {year} {2016})}\BibitemShut
  {NoStop}%
\bibitem [{\citenamefont {Wang}\ \emph {et~al.}(2019)\citenamefont {Wang},
  \citenamefont {Wieder}, \citenamefont {Li}, \citenamefont {Yan},\ and\
  \citenamefont {Bernevig}}]{wang2019higher}%
  \BibitemOpen
  \bibfield  {author} {\bibinfo {author} {\bibfnamefont {Z.}~\bibnamefont
  {Wang}}, \bibinfo {author} {\bibfnamefont {B.~J.}\ \bibnamefont {Wieder}},
  \bibinfo {author} {\bibfnamefont {J.}~\bibnamefont {Li}}, \bibinfo {author}
  {\bibfnamefont {B.}~\bibnamefont {Yan}},\ and\ \bibinfo {author}
  {\bibfnamefont {B.~A.}\ \bibnamefont {Bernevig}},\ }\href@noop {} {\bibfield
  {journal} {\bibinfo  {journal} {Physical review letters}\ }\textbf {\bibinfo
  {volume} {123}},\ \bibinfo {pages} {186401} (\bibinfo {year}
  {2019})}\BibitemShut {NoStop}%
\bibitem [{\citenamefont {Tamai}\ \emph {et~al.}(2016)\citenamefont {Tamai},
  \citenamefont {Wu}, \citenamefont {Cucchi}, \citenamefont {Bruno},
  \citenamefont {Ricc{\`o}}, \citenamefont {Kim}, \citenamefont {Hoesch},
  \citenamefont {Barreteau}, \citenamefont {Giannini}, \citenamefont {Besnard}
  \emph {et~al.}}]{tamai2016fermi}%
  \BibitemOpen
  \bibfield  {author} {\bibinfo {author} {\bibfnamefont {A.}~\bibnamefont
  {Tamai}}, \bibinfo {author} {\bibfnamefont {Q.}~\bibnamefont {Wu}}, \bibinfo
  {author} {\bibfnamefont {I.}~\bibnamefont {Cucchi}}, \bibinfo {author}
  {\bibfnamefont {F.~Y.}\ \bibnamefont {Bruno}}, \bibinfo {author}
  {\bibfnamefont {S.}~\bibnamefont {Ricc{\`o}}}, \bibinfo {author}
  {\bibfnamefont {T.~K.}\ \bibnamefont {Kim}}, \bibinfo {author} {\bibfnamefont
  {M.}~\bibnamefont {Hoesch}}, \bibinfo {author} {\bibfnamefont
  {C.}~\bibnamefont {Barreteau}}, \bibinfo {author} {\bibfnamefont
  {E.}~\bibnamefont {Giannini}}, \bibinfo {author} {\bibfnamefont
  {C.}~\bibnamefont {Besnard}}, \emph {et~al.},\ }\href@noop {} {\bibfield
  {journal} {\bibinfo  {journal} {Physical Review X}\ }\textbf {\bibinfo
  {volume} {6}},\ \bibinfo {pages} {031021} (\bibinfo {year}
  {2016})}\BibitemShut {NoStop}%
\bibitem [{\citenamefont {Xu}\ \emph {et~al.}(2018)\citenamefont {Xu},
  \citenamefont {Wang}, \citenamefont {Magrez}, \citenamefont {Bugnon},
  \citenamefont {Berger}, \citenamefont {Matt}, \citenamefont {Strocov},
  \citenamefont {Plumb}, \citenamefont {Radovic}, \citenamefont {Pomjakushina}
  \emph {et~al.}}]{xu2018evidence}%
  \BibitemOpen
  \bibfield  {author} {\bibinfo {author} {\bibfnamefont {N.}~\bibnamefont
  {Xu}}, \bibinfo {author} {\bibfnamefont {Z.}~\bibnamefont {Wang}}, \bibinfo
  {author} {\bibfnamefont {A.}~\bibnamefont {Magrez}}, \bibinfo {author}
  {\bibfnamefont {P.}~\bibnamefont {Bugnon}}, \bibinfo {author} {\bibfnamefont
  {H.}~\bibnamefont {Berger}}, \bibinfo {author} {\bibfnamefont {C.~E.}\
  \bibnamefont {Matt}}, \bibinfo {author} {\bibfnamefont {V.~N.}\ \bibnamefont
  {Strocov}}, \bibinfo {author} {\bibfnamefont {N.~C.}\ \bibnamefont {Plumb}},
  \bibinfo {author} {\bibfnamefont {M.}~\bibnamefont {Radovic}}, \bibinfo
  {author} {\bibfnamefont {E.}~\bibnamefont {Pomjakushina}}, \emph {et~al.},\
  }\href@noop {} {\bibfield  {journal} {\bibinfo  {journal} {Physical Review
  Letters}\ }\textbf {\bibinfo {volume} {121}},\ \bibinfo {pages} {136401}
  (\bibinfo {year} {2018})}\BibitemShut {NoStop}%
\bibitem [{\citenamefont {Lu}\ \emph {et~al.}(2020)\citenamefont {Lu},
  \citenamefont {Sapkota}, \citenamefont {DeBeer-Schmitt}, \citenamefont {Wu},
  \citenamefont {Cao}, \citenamefont {Mannella}, \citenamefont {Mandrus},
  \citenamefont {Aczel},\ and\ \citenamefont {MacDougall}}]{lu2020canted}%
  \BibitemOpen
  \bibfield  {author} {\bibinfo {author} {\bibfnamefont {K.}~\bibnamefont
  {Lu}}, \bibinfo {author} {\bibfnamefont {D.}~\bibnamefont {Sapkota}},
  \bibinfo {author} {\bibfnamefont {L.}~\bibnamefont {DeBeer-Schmitt}},
  \bibinfo {author} {\bibfnamefont {Y.}~\bibnamefont {Wu}}, \bibinfo {author}
  {\bibfnamefont {H.}~\bibnamefont {Cao}}, \bibinfo {author} {\bibfnamefont
  {N.}~\bibnamefont {Mannella}}, \bibinfo {author} {\bibfnamefont
  {D.}~\bibnamefont {Mandrus}}, \bibinfo {author} {\bibfnamefont {A.~A.}\
  \bibnamefont {Aczel}},\ and\ \bibinfo {author} {\bibfnamefont {G.~J.}\
  \bibnamefont {MacDougall}},\ }\href@noop {} {\bibfield  {journal} {\bibinfo
  {journal} {Physical Review Materials}\ }\textbf {\bibinfo {volume} {4}},\
  \bibinfo {pages} {054416} (\bibinfo {year} {2020})}\BibitemShut {NoStop}%
\bibitem [{\citenamefont {Kousaka}\ \emph {et~al.}(2016)\citenamefont
  {Kousaka}, \citenamefont {Ogura}, \citenamefont {Zhang}, \citenamefont
  {Miao}, \citenamefont {Lee}, \citenamefont {Torii}, \citenamefont {Kamiyama},
  \citenamefont {Campo}, \citenamefont {Inoue},\ and\ \citenamefont
  {Akimitsu}}]{kousaka2016long}%
  \BibitemOpen
  \bibfield  {author} {\bibinfo {author} {\bibfnamefont {Y.}~\bibnamefont
  {Kousaka}}, \bibinfo {author} {\bibfnamefont {T.}~\bibnamefont {Ogura}},
  \bibinfo {author} {\bibfnamefont {J.}~\bibnamefont {Zhang}}, \bibinfo
  {author} {\bibfnamefont {P.}~\bibnamefont {Miao}}, \bibinfo {author}
  {\bibfnamefont {S.}~\bibnamefont {Lee}}, \bibinfo {author} {\bibfnamefont
  {S.}~\bibnamefont {Torii}}, \bibinfo {author} {\bibfnamefont
  {T.}~\bibnamefont {Kamiyama}}, \bibinfo {author} {\bibfnamefont
  {J.}~\bibnamefont {Campo}}, \bibinfo {author} {\bibfnamefont
  {K.}~\bibnamefont {Inoue}},\ and\ \bibinfo {author} {\bibfnamefont
  {J.}~\bibnamefont {Akimitsu}},\ }in\ \href@noop {} {\emph {\bibinfo
  {booktitle} {Journal of Physics: Conference Series}}},\ Vol.\ \bibinfo
  {volume} {746}\ (\bibinfo {organization} {IOP Publishing},\ \bibinfo {year}
  {2016})\ p.\ \bibinfo {pages} {012061}\BibitemShut {NoStop}%
\bibitem [{\citenamefont {Zhang}\ \emph {et~al.}(2018)\citenamefont {Zhang},
  \citenamefont {Wei}, \citenamefont {Zheng}, \citenamefont {Lu}, \citenamefont
  {Wu}, \citenamefont {Zhu}, \citenamefont {Tang}, \citenamefont {Ning},
  \citenamefont {Han}, \citenamefont {Ling} \emph
  {et~al.}}]{zhang2018electrical}%
  \BibitemOpen
  \bibfield  {author} {\bibinfo {author} {\bibfnamefont {H.}~\bibnamefont
  {Zhang}}, \bibinfo {author} {\bibfnamefont {W.}~\bibnamefont {Wei}}, \bibinfo
  {author} {\bibfnamefont {G.}~\bibnamefont {Zheng}}, \bibinfo {author}
  {\bibfnamefont {J.}~\bibnamefont {Lu}}, \bibinfo {author} {\bibfnamefont
  {M.}~\bibnamefont {Wu}}, \bibinfo {author} {\bibfnamefont {X.}~\bibnamefont
  {Zhu}}, \bibinfo {author} {\bibfnamefont {J.}~\bibnamefont {Tang}}, \bibinfo
  {author} {\bibfnamefont {W.}~\bibnamefont {Ning}}, \bibinfo {author}
  {\bibfnamefont {Y.}~\bibnamefont {Han}}, \bibinfo {author} {\bibfnamefont
  {L.}~\bibnamefont {Ling}}, \emph {et~al.},\ }\href@noop {} {\bibfield
  {journal} {\bibinfo  {journal} {Applied Physics Letters}\ }\textbf {\bibinfo
  {volume} {113}} (\bibinfo {year} {2018})}\BibitemShut {NoStop}%
\bibitem [{\citenamefont {Parkin}\ \emph {et~al.}(1983)\citenamefont {Parkin},
  \citenamefont {Marseglia},\ and\ \citenamefont {Brown}}]{parkin1983magnetic}%
  \BibitemOpen
  \bibfield  {author} {\bibinfo {author} {\bibfnamefont {S.}~\bibnamefont
  {Parkin}}, \bibinfo {author} {\bibfnamefont {E.}~\bibnamefont {Marseglia}},\
  and\ \bibinfo {author} {\bibfnamefont {P.}~\bibnamefont {Brown}},\
  }\href@noop {} {\bibfield  {journal} {\bibinfo  {journal} {Journal of Physics
  C: Solid State Physics}\ }\textbf {\bibinfo {volume} {16}},\ \bibinfo {pages}
  {2765} (\bibinfo {year} {1983})}\BibitemShut {NoStop}%
\bibitem [{\citenamefont {Parkin}\ and\ \citenamefont
  {Friend}(1980)}]{parkin19803}%
  \BibitemOpen
  \bibfield  {author} {\bibinfo {author} {\bibfnamefont {S.}~\bibnamefont
  {Parkin}}\ and\ \bibinfo {author} {\bibfnamefont {R.}~\bibnamefont
  {Friend}},\ }\href@noop {} {\bibfield  {journal} {\bibinfo  {journal}
  {Philosophical Magazine B}\ }\textbf {\bibinfo {volume} {41}},\ \bibinfo
  {pages} {65} (\bibinfo {year} {1980})}\BibitemShut {NoStop}%
\bibitem [{\citenamefont {Morosan}\ \emph {et~al.}(2007)\citenamefont
  {Morosan}, \citenamefont {Zandbergen}, \citenamefont {Li}, \citenamefont
  {Lee}, \citenamefont {Checkelsky}, \citenamefont {Heinrich}, \citenamefont
  {Siegrist}, \citenamefont {Ong},\ and\ \citenamefont
  {Cava}}]{morosan2007sharp}%
  \BibitemOpen
  \bibfield  {author} {\bibinfo {author} {\bibfnamefont {E.}~\bibnamefont
  {Morosan}}, \bibinfo {author} {\bibfnamefont {H.}~\bibnamefont {Zandbergen}},
  \bibinfo {author} {\bibfnamefont {L.}~\bibnamefont {Li}}, \bibinfo {author}
  {\bibfnamefont {M.}~\bibnamefont {Lee}}, \bibinfo {author} {\bibfnamefont
  {J.}~\bibnamefont {Checkelsky}}, \bibinfo {author} {\bibfnamefont
  {M.}~\bibnamefont {Heinrich}}, \bibinfo {author} {\bibfnamefont
  {T.}~\bibnamefont {Siegrist}}, \bibinfo {author} {\bibfnamefont {N.~P.}\
  \bibnamefont {Ong}},\ and\ \bibinfo {author} {\bibfnamefont {R.}~\bibnamefont
  {Cava}},\ }\href@noop {} {\bibfield  {journal} {\bibinfo  {journal} {Physical
  Review B}\ }\textbf {\bibinfo {volume} {75}},\ \bibinfo {pages} {104401}
  (\bibinfo {year} {2007})}\BibitemShut {NoStop}%
\bibitem [{\citenamefont {Checkelsky}\ \emph {et~al.}(2008)\citenamefont
  {Checkelsky}, \citenamefont {Lee}, \citenamefont {Morosan}, \citenamefont
  {Cava},\ and\ \citenamefont {Ong}}]{checkelsky2008anomalous}%
  \BibitemOpen
  \bibfield  {author} {\bibinfo {author} {\bibfnamefont {J.}~\bibnamefont
  {Checkelsky}}, \bibinfo {author} {\bibfnamefont {M.}~\bibnamefont {Lee}},
  \bibinfo {author} {\bibfnamefont {E.}~\bibnamefont {Morosan}}, \bibinfo
  {author} {\bibfnamefont {R.}~\bibnamefont {Cava}},\ and\ \bibinfo {author}
  {\bibfnamefont {N.}~\bibnamefont {Ong}},\ }\href@noop {} {\bibfield
  {journal} {\bibinfo  {journal} {Physical Review B}\ }\textbf {\bibinfo
  {volume} {77}},\ \bibinfo {pages} {014433} (\bibinfo {year}
  {2008})}\BibitemShut {NoStop}%
\bibitem [{\citenamefont {Dijkstra}\ \emph {et~al.}(1989)\citenamefont
  {Dijkstra}, \citenamefont {Zijlema}, \citenamefont {Van~Bruggen},
  \citenamefont {Haas},\ and\ \citenamefont {De~Groot}}]{dijkstra1989band}%
  \BibitemOpen
  \bibfield  {author} {\bibinfo {author} {\bibfnamefont {J.}~\bibnamefont
  {Dijkstra}}, \bibinfo {author} {\bibfnamefont {P.}~\bibnamefont {Zijlema}},
  \bibinfo {author} {\bibfnamefont {C.}~\bibnamefont {Van~Bruggen}}, \bibinfo
  {author} {\bibfnamefont {C.}~\bibnamefont {Haas}},\ and\ \bibinfo {author}
  {\bibfnamefont {R.}~\bibnamefont {De~Groot}},\ }\href@noop {} {\bibfield
  {journal} {\bibinfo  {journal} {Journal of Physics: Condensed Matter}\
  }\textbf {\bibinfo {volume} {1}},\ \bibinfo {pages} {6363} (\bibinfo {year}
  {1989})}\BibitemShut {NoStop}%
\bibitem [{\citenamefont {Hardy}\ \emph {et~al.}(2015)\citenamefont {Hardy},
  \citenamefont {Chen}, \citenamefont {Marcinkova}, \citenamefont {Ji},
  \citenamefont {Sinova}, \citenamefont {Natelson},\ and\ \citenamefont
  {Morosan}}]{hardy2015very}%
  \BibitemOpen
  \bibfield  {author} {\bibinfo {author} {\bibfnamefont {W.~J.}\ \bibnamefont
  {Hardy}}, \bibinfo {author} {\bibfnamefont {C.-W.}\ \bibnamefont {Chen}},
  \bibinfo {author} {\bibfnamefont {A.}~\bibnamefont {Marcinkova}}, \bibinfo
  {author} {\bibfnamefont {H.}~\bibnamefont {Ji}}, \bibinfo {author}
  {\bibfnamefont {J.}~\bibnamefont {Sinova}}, \bibinfo {author} {\bibfnamefont
  {D.}~\bibnamefont {Natelson}},\ and\ \bibinfo {author} {\bibfnamefont
  {E.}~\bibnamefont {Morosan}},\ }\href@noop {} {\bibfield  {journal} {\bibinfo
   {journal} {Physical Review B}\ }\textbf {\bibinfo {volume} {91}},\ \bibinfo
  {pages} {054426} (\bibinfo {year} {2015})}\BibitemShut {NoStop}%
\bibitem [{\citenamefont {Narita}\ \emph {et~al.}(1994)\citenamefont {Narita},
  \citenamefont {Ikuta}, \citenamefont {Hinode}, \citenamefont {Uchida},
  \citenamefont {Ohtani},\ and\ \citenamefont
  {Wakihara}}]{narita1994preparation}%
  \BibitemOpen
  \bibfield  {author} {\bibinfo {author} {\bibfnamefont {H.}~\bibnamefont
  {Narita}}, \bibinfo {author} {\bibfnamefont {H.}~\bibnamefont {Ikuta}},
  \bibinfo {author} {\bibfnamefont {H.}~\bibnamefont {Hinode}}, \bibinfo
  {author} {\bibfnamefont {T.}~\bibnamefont {Uchida}}, \bibinfo {author}
  {\bibfnamefont {T.}~\bibnamefont {Ohtani}},\ and\ \bibinfo {author}
  {\bibfnamefont {M.}~\bibnamefont {Wakihara}},\ }\href@noop {} {\bibfield
  {journal} {\bibinfo  {journal} {Journal of Solid State Chemistry}\ }\textbf
  {\bibinfo {volume} {108}},\ \bibinfo {pages} {148} (\bibinfo {year}
  {1994})}\BibitemShut {NoStop}%
\bibitem [{\citenamefont {Yang}\ \emph {et~al.}(2021)\citenamefont {Yang},
  \citenamefont {Yu}, \citenamefont {Xu}, \citenamefont {Wang}, \citenamefont
  {Hu}, \citenamefont {Xu}, \citenamefont {Wang}, \citenamefont {Zhang},
  \citenamefont {Ren}, \citenamefont {Xu} \emph
  {et~al.}}]{yang2021anisotropic}%
  \BibitemOpen
  \bibfield  {author} {\bibinfo {author} {\bibfnamefont {X.}~\bibnamefont
  {Yang}}, \bibinfo {author} {\bibfnamefont {T.}~\bibnamefont {Yu}}, \bibinfo
  {author} {\bibfnamefont {C.}~\bibnamefont {Xu}}, \bibinfo {author}
  {\bibfnamefont {J.}~\bibnamefont {Wang}}, \bibinfo {author} {\bibfnamefont
  {W.}~\bibnamefont {Hu}}, \bibinfo {author} {\bibfnamefont {Z.}~\bibnamefont
  {Xu}}, \bibinfo {author} {\bibfnamefont {T.}~\bibnamefont {Wang}}, \bibinfo
  {author} {\bibfnamefont {C.}~\bibnamefont {Zhang}}, \bibinfo {author}
  {\bibfnamefont {Z.}~\bibnamefont {Ren}}, \bibinfo {author} {\bibfnamefont
  {Z.-a.}\ \bibnamefont {Xu}}, \emph {et~al.},\ }\href@noop {} {\bibfield
  {journal} {\bibinfo  {journal} {Physical Review B}\ }\textbf {\bibinfo
  {volume} {104}},\ \bibinfo {pages} {035157} (\bibinfo {year}
  {2021})}\BibitemShut {NoStop}%
\bibitem [{\citenamefont {Park}\ \emph {et~al.}(2022)\citenamefont {Park},
  \citenamefont {Kang}, \citenamefont {Kim}, \citenamefont {Lee}, \citenamefont
  {Noh}, \citenamefont {Han},\ and\ \citenamefont {Park}}]{park2022field}%
  \BibitemOpen
  \bibfield  {author} {\bibinfo {author} {\bibfnamefont {P.}~\bibnamefont
  {Park}}, \bibinfo {author} {\bibfnamefont {Y.-G.}\ \bibnamefont {Kang}},
  \bibinfo {author} {\bibfnamefont {J.}~\bibnamefont {Kim}}, \bibinfo {author}
  {\bibfnamefont {K.~H.}\ \bibnamefont {Lee}}, \bibinfo {author} {\bibfnamefont
  {H.-J.}\ \bibnamefont {Noh}}, \bibinfo {author} {\bibfnamefont {M.~J.}\
  \bibnamefont {Han}},\ and\ \bibinfo {author} {\bibfnamefont {J.-G.}\
  \bibnamefont {Park}},\ }\href@noop {} {\bibfield  {journal} {\bibinfo
  {journal} {npj Quantum Materials}\ }\textbf {\bibinfo {volume} {7}},\
  \bibinfo {pages} {42} (\bibinfo {year} {2022})}\BibitemShut {NoStop}%
\bibitem [{\citenamefont {Liu}\ \emph {et~al.}(2021)\citenamefont {Liu},
  \citenamefont {Hu}, \citenamefont {Stavitski}, \citenamefont {Attenkofer},
  \citenamefont {Petrovic} \emph {et~al.}}]{liu2021magnetic}%
  \BibitemOpen
  \bibfield  {author} {\bibinfo {author} {\bibfnamefont {Y.}~\bibnamefont
  {Liu}}, \bibinfo {author} {\bibfnamefont {Z.}~\bibnamefont {Hu}}, \bibinfo
  {author} {\bibfnamefont {E.}~\bibnamefont {Stavitski}}, \bibinfo {author}
  {\bibfnamefont {K.}~\bibnamefont {Attenkofer}}, \bibinfo {author}
  {\bibfnamefont {C.}~\bibnamefont {Petrovic}}, \emph {et~al.},\ }\href@noop {}
  {\bibfield  {journal} {\bibinfo  {journal} {Physical Review Research}\
  }\textbf {\bibinfo {volume} {3}},\ \bibinfo {pages} {023181} (\bibinfo {year}
  {2021})}\BibitemShut {NoStop}%
\bibitem [{\citenamefont {Liu}\ \emph {et~al.}(2022)\citenamefont {Liu},
  \citenamefont {Hu}, \citenamefont {Tong}, \citenamefont {Bauer},
  \citenamefont {Petrovic} \emph {et~al.}}]{liu2022electrical}%
  \BibitemOpen
  \bibfield  {author} {\bibinfo {author} {\bibfnamefont {Y.}~\bibnamefont
  {Liu}}, \bibinfo {author} {\bibfnamefont {Z.}~\bibnamefont {Hu}}, \bibinfo
  {author} {\bibfnamefont {X.}~\bibnamefont {Tong}}, \bibinfo {author}
  {\bibfnamefont {E.~D.}\ \bibnamefont {Bauer}}, \bibinfo {author}
  {\bibfnamefont {C.}~\bibnamefont {Petrovic}}, \emph {et~al.},\ }\href@noop {}
  {\bibfield  {journal} {\bibinfo  {journal} {Physical Review Research}\
  }\textbf {\bibinfo {volume} {4}},\ \bibinfo {pages} {013048} (\bibinfo {year}
  {2022})}\BibitemShut {NoStop}%
\bibitem [{\citenamefont {Algaidi}\ \emph {et~al.}(2023)\citenamefont
  {Algaidi}, \citenamefont {Zhang}, \citenamefont {Liu}, \citenamefont {Zheng},
  \citenamefont {Ma}, \citenamefont {Yuan},\ and\ \citenamefont
  {Zhang}}]{algaidi2023thickness}%
  \BibitemOpen
  \bibfield  {author} {\bibinfo {author} {\bibfnamefont {H.}~\bibnamefont
  {Algaidi}}, \bibinfo {author} {\bibfnamefont {C.}~\bibnamefont {Zhang}},
  \bibinfo {author} {\bibfnamefont {C.}~\bibnamefont {Liu}}, \bibinfo {author}
  {\bibfnamefont {D.}~\bibnamefont {Zheng}}, \bibinfo {author} {\bibfnamefont
  {Y.}~\bibnamefont {Ma}}, \bibinfo {author} {\bibfnamefont {Y.}~\bibnamefont
  {Yuan}},\ and\ \bibinfo {author} {\bibfnamefont {X.}~\bibnamefont {Zhang}},\
  }\href@noop {} {\bibfield  {journal} {\bibinfo  {journal} {Physical Review
  B}\ }\textbf {\bibinfo {volume} {107}},\ \bibinfo {pages} {134406} (\bibinfo
  {year} {2023})}\BibitemShut {NoStop}%
\bibitem [{\citenamefont {Nozieres}\ and\ \citenamefont
  {Blandin}(1980)}]{nozieres1980kondo}%
  \BibitemOpen
  \bibfield  {author} {\bibinfo {author} {\bibfnamefont {P.}~\bibnamefont
  {Nozieres}}\ and\ \bibinfo {author} {\bibfnamefont {A.}~\bibnamefont
  {Blandin}},\ }\href@noop {} {\bibfield  {journal} {\bibinfo  {journal}
  {Journal de Physique}\ }\textbf {\bibinfo {volume} {41}},\ \bibinfo {pages}
  {193} (\bibinfo {year} {1980})}\BibitemShut {NoStop}%
\bibitem [{\citenamefont {Van~Laar}\ \emph {et~al.}(1971)\citenamefont
  {Van~Laar}, \citenamefont {Rietveld},\ and\ \citenamefont
  {Ijdo}}]{van1971magnetic}%
  \BibitemOpen
  \bibfield  {author} {\bibinfo {author} {\bibfnamefont {B.}~\bibnamefont
  {Van~Laar}}, \bibinfo {author} {\bibfnamefont {H.}~\bibnamefont {Rietveld}},\
  and\ \bibinfo {author} {\bibfnamefont {D.}~\bibnamefont {Ijdo}},\ }\href@noop
  {} {\bibfield  {journal} {\bibinfo  {journal} {Journal of Solid State
  Chemistry}\ }\textbf {\bibinfo {volume} {3}},\ \bibinfo {pages} {154}
  (\bibinfo {year} {1971})}\BibitemShut {NoStop}%
\bibitem [{\citenamefont {Dragomir}\ \emph {et~al.}(2019)\citenamefont
  {Dragomir}, \citenamefont {Dube}, \citenamefont {Arcon}, \citenamefont
  {Boyer}, \citenamefont {Rutherford}, \citenamefont {Wiebe}, \citenamefont
  {King}, \citenamefont {Dabkowska},\ and\ \citenamefont
  {Greedan}}]{dragomir2019comparing}%
  \BibitemOpen
  \bibfield  {author} {\bibinfo {author} {\bibfnamefont {M.}~\bibnamefont
  {Dragomir}}, \bibinfo {author} {\bibfnamefont {P.~A.}\ \bibnamefont {Dube}},
  \bibinfo {author} {\bibfnamefont {I.}~\bibnamefont {Arcon}}, \bibinfo
  {author} {\bibfnamefont {C.}~\bibnamefont {Boyer}}, \bibinfo {author}
  {\bibfnamefont {M.}~\bibnamefont {Rutherford}}, \bibinfo {author}
  {\bibfnamefont {C.~R.}\ \bibnamefont {Wiebe}}, \bibinfo {author}
  {\bibfnamefont {G.}~\bibnamefont {King}}, \bibinfo {author} {\bibfnamefont
  {H.~A.}\ \bibnamefont {Dabkowska}},\ and\ \bibinfo {author} {\bibfnamefont
  {J.~E.}\ \bibnamefont {Greedan}},\ }\href@noop {} {\bibfield  {journal}
  {\bibinfo  {journal} {Chemistry of materials}\ }\textbf {\bibinfo {volume}
  {31}},\ \bibinfo {pages} {7833} (\bibinfo {year} {2019})}\BibitemShut
  {NoStop}%
\bibitem [{\citenamefont {Park}\ \emph {et~al.}(2024)\citenamefont {Park},
  \citenamefont {Cho}, \citenamefont {Kim}, \citenamefont {An}, \citenamefont
  {Avdeev}, \citenamefont {Iida}, \citenamefont {Kajimoto},\ and\ \citenamefont
  {Park}}]{park2024composition}%
  \BibitemOpen
  \bibfield  {author} {\bibinfo {author} {\bibfnamefont {P.}~\bibnamefont
  {Park}}, \bibinfo {author} {\bibfnamefont {W.}~\bibnamefont {Cho}}, \bibinfo
  {author} {\bibfnamefont {C.}~\bibnamefont {Kim}}, \bibinfo {author}
  {\bibfnamefont {Y.}~\bibnamefont {An}}, \bibinfo {author} {\bibfnamefont
  {M.}~\bibnamefont {Avdeev}}, \bibinfo {author} {\bibfnamefont
  {K.}~\bibnamefont {Iida}}, \bibinfo {author} {\bibfnamefont {R.}~\bibnamefont
  {Kajimoto}},\ and\ \bibinfo {author} {\bibfnamefont {J.-G.}\ \bibnamefont
  {Park}},\ }\href@noop {} {\bibfield  {journal} {\bibinfo  {journal} {Physical
  Review B}\ }\textbf {\bibinfo {volume} {109}},\ \bibinfo {pages} {L060403}
  (\bibinfo {year} {2024})}\BibitemShut {NoStop}%
\bibitem [{\citenamefont {Park}\ \emph {et~al.}(2023)\citenamefont {Park},
  \citenamefont {Cho}, \citenamefont {Kim}, \citenamefont {An}, \citenamefont
  {Kang}, \citenamefont {Avdeev}, \citenamefont {Sibille}, \citenamefont
  {Iida}, \citenamefont {Kajimoto}, \citenamefont {Lee} \emph
  {et~al.}}]{park2023tetrahedral}%
  \BibitemOpen
  \bibfield  {author} {\bibinfo {author} {\bibfnamefont {P.}~\bibnamefont
  {Park}}, \bibinfo {author} {\bibfnamefont {W.}~\bibnamefont {Cho}}, \bibinfo
  {author} {\bibfnamefont {C.}~\bibnamefont {Kim}}, \bibinfo {author}
  {\bibfnamefont {Y.}~\bibnamefont {An}}, \bibinfo {author} {\bibfnamefont
  {Y.-G.}\ \bibnamefont {Kang}}, \bibinfo {author} {\bibfnamefont
  {M.}~\bibnamefont {Avdeev}}, \bibinfo {author} {\bibfnamefont
  {R.}~\bibnamefont {Sibille}}, \bibinfo {author} {\bibfnamefont
  {K.}~\bibnamefont {Iida}}, \bibinfo {author} {\bibfnamefont {R.}~\bibnamefont
  {Kajimoto}}, \bibinfo {author} {\bibfnamefont {K.~H.}\ \bibnamefont {Lee}},
  \emph {et~al.},\ }\href@noop {} {\bibfield  {journal} {\bibinfo  {journal}
  {Nature Communications}\ }\textbf {\bibinfo {volume} {14}},\ \bibinfo {pages}
  {8346} (\bibinfo {year} {2023})}\BibitemShut {NoStop}%
\bibitem [{\citenamefont {Aristov}(1997)}]{aristov1997indirect}%
  \BibitemOpen
  \bibfield  {author} {\bibinfo {author} {\bibfnamefont {D.}~\bibnamefont
  {Aristov}},\ }\href@noop {} {\bibfield  {journal} {\bibinfo  {journal}
  {Physical Review B}\ }\textbf {\bibinfo {volume} {55}},\ \bibinfo {pages}
  {8064} (\bibinfo {year} {1997})}\BibitemShut {NoStop}%
\bibitem [{\citenamefont {Yosida}(1957)}]{yosida1957magnetic}%
  \BibitemOpen
  \bibfield  {author} {\bibinfo {author} {\bibfnamefont {K.}~\bibnamefont
  {Yosida}},\ }\href@noop {} {\bibfield  {journal} {\bibinfo  {journal}
  {Physical Review}\ }\textbf {\bibinfo {volume} {106}},\ \bibinfo {pages}
  {893} (\bibinfo {year} {1957})}\BibitemShut {NoStop}%
\bibitem [{\citenamefont {Hewson}(1997)}]{hewson1997kondo}%
  \BibitemOpen
  \bibfield  {author} {\bibinfo {author} {\bibfnamefont {A.~C.}\ \bibnamefont
  {Hewson}},\ }\href@noop {} {\emph {\bibinfo {title} {The Kondo problem to
  heavy fermions}}},\ \bibinfo {number} {2}\ (\bibinfo  {publisher} {Cambridge
  university press},\ \bibinfo {year} {1997})\BibitemShut {NoStop}%
\bibitem [{\citenamefont {Kondo}(1964)}]{kondo1964resistance}%
  \BibitemOpen
  \bibfield  {author} {\bibinfo {author} {\bibfnamefont {J.}~\bibnamefont
  {Kondo}},\ }\href@noop {} {\bibfield  {journal} {\bibinfo  {journal}
  {Progress of theoretical physics}\ }\textbf {\bibinfo {volume} {32}},\
  \bibinfo {pages} {37} (\bibinfo {year} {1964})}\BibitemShut {NoStop}%
\bibitem [{\citenamefont {Kondo}(1970)}]{kondo1970theory}%
  \BibitemOpen
  \bibfield  {author} {\bibinfo {author} {\bibfnamefont {J.}~\bibnamefont
  {Kondo}},\ }in\ \href@noop {} {\emph {\bibinfo {booktitle} {Solid state
  physics}}},\ Vol.~\bibinfo {volume} {23}\ (\bibinfo  {publisher} {Elsevier},\
  \bibinfo {year} {1970})\ pp.\ \bibinfo {pages} {183--281}\BibitemShut
  {NoStop}%
\bibitem [{\citenamefont {Nagai}\ and\ \citenamefont
  {Kondo}(1975)}]{nagai1975resistivity}%
  \BibitemOpen
  \bibfield  {author} {\bibinfo {author} {\bibfnamefont {S.}~\bibnamefont
  {Nagai}}\ and\ \bibinfo {author} {\bibfnamefont {J.}~\bibnamefont {Kondo}},\
  }\href@noop {} {\bibfield  {journal} {\bibinfo  {journal} {Journal of the
  Physical Society of Japan}\ }\textbf {\bibinfo {volume} {38}},\ \bibinfo
  {pages} {129} (\bibinfo {year} {1975})}\BibitemShut {NoStop}%
\bibitem [{\citenamefont {Altshuler}\ and\ \citenamefont
  {Aronov}(1985)}]{altshuler1985electron}%
  \BibitemOpen
  \bibfield  {author} {\bibinfo {author} {\bibfnamefont {B.~L.}\ \bibnamefont
  {Altshuler}}\ and\ \bibinfo {author} {\bibfnamefont {A.~G.}\ \bibnamefont
  {Aronov}},\ }in\ \href@noop {} {\emph {\bibinfo {booktitle} {Modern Problems
  in condensed matter sciences}}},\ Vol.~\bibinfo {volume} {10}\ (\bibinfo
  {publisher} {Elsevier},\ \bibinfo {year} {1985})\ pp.\ \bibinfo {pages}
  {1--153}\BibitemShut {NoStop}%
\bibitem [{\citenamefont {Lee}\ and\ \citenamefont
  {Ramakrishnan}(1985)}]{lee1985disordered}%
  \BibitemOpen
  \bibfield  {author} {\bibinfo {author} {\bibfnamefont {P.~A.}\ \bibnamefont
  {Lee}}\ and\ \bibinfo {author} {\bibfnamefont {T.}~\bibnamefont
  {Ramakrishnan}},\ }\href@noop {} {\bibfield  {journal} {\bibinfo  {journal}
  {Reviews of modern physics}\ }\textbf {\bibinfo {volume} {57}},\ \bibinfo
  {pages} {287} (\bibinfo {year} {1985})}\BibitemShut {NoStop}%
\bibitem [{\citenamefont {Cichorek}\ \emph {et~al.}(2005)\citenamefont
  {Cichorek}, \citenamefont {Sanchez}, \citenamefont {Gegenwart}, \citenamefont
  {Weickert}, \citenamefont {Wojakowski}, \citenamefont {Henkie}, \citenamefont
  {Auffermann}, \citenamefont {Paschen}, \citenamefont {Kniep},\ and\
  \citenamefont {Steglich}}]{cichorek2005two}%
  \BibitemOpen
  \bibfield  {author} {\bibinfo {author} {\bibfnamefont {T.}~\bibnamefont
  {Cichorek}}, \bibinfo {author} {\bibfnamefont {A.}~\bibnamefont {Sanchez}},
  \bibinfo {author} {\bibfnamefont {P.}~\bibnamefont {Gegenwart}}, \bibinfo
  {author} {\bibfnamefont {F.}~\bibnamefont {Weickert}}, \bibinfo {author}
  {\bibfnamefont {A.}~\bibnamefont {Wojakowski}}, \bibinfo {author}
  {\bibfnamefont {Z.}~\bibnamefont {Henkie}}, \bibinfo {author} {\bibfnamefont
  {G.}~\bibnamefont {Auffermann}}, \bibinfo {author} {\bibfnamefont
  {S.}~\bibnamefont {Paschen}}, \bibinfo {author} {\bibfnamefont
  {R.}~\bibnamefont {Kniep}},\ and\ \bibinfo {author} {\bibfnamefont
  {F.}~\bibnamefont {Steglich}},\ }\href@noop {} {\bibfield  {journal}
  {\bibinfo  {journal} {Physical review letters}\ }\textbf {\bibinfo {volume}
  {94}},\ \bibinfo {pages} {236603} (\bibinfo {year} {2005})}\BibitemShut
  {NoStop}%
\bibitem [{\citenamefont {Cichorek}\ \emph {et~al.}(2016)\citenamefont
  {Cichorek}, \citenamefont {Bochenek}, \citenamefont {Schmidt}, \citenamefont
  {Czulucki}, \citenamefont {Auffermann}, \citenamefont {Kniep}, \citenamefont
  {Niewa}, \citenamefont {Steglich},\ and\ \citenamefont
  {Kirchner}}]{cichorek2016two}%
  \BibitemOpen
  \bibfield  {author} {\bibinfo {author} {\bibfnamefont {T.}~\bibnamefont
  {Cichorek}}, \bibinfo {author} {\bibfnamefont {L.}~\bibnamefont {Bochenek}},
  \bibinfo {author} {\bibfnamefont {M.}~\bibnamefont {Schmidt}}, \bibinfo
  {author} {\bibfnamefont {A.}~\bibnamefont {Czulucki}}, \bibinfo {author}
  {\bibfnamefont {G.}~\bibnamefont {Auffermann}}, \bibinfo {author}
  {\bibfnamefont {R.}~\bibnamefont {Kniep}}, \bibinfo {author} {\bibfnamefont
  {R.}~\bibnamefont {Niewa}}, \bibinfo {author} {\bibfnamefont
  {F.}~\bibnamefont {Steglich}},\ and\ \bibinfo {author} {\bibfnamefont
  {S.}~\bibnamefont {Kirchner}},\ }\href@noop {} {\bibfield  {journal}
  {\bibinfo  {journal} {Physical Review Letters}\ }\textbf {\bibinfo {volume}
  {117}},\ \bibinfo {pages} {106601} (\bibinfo {year} {2016})}\BibitemShut
  {NoStop}%
\bibitem [{\citenamefont {Zhu}\ \emph {et~al.}(2016{\natexlab{a}})\citenamefont
  {Zhu}, \citenamefont {Nie}, \citenamefont {Xiong}, \citenamefont
  {Schlottmann},\ and\ \citenamefont {Zhao}}]{zhu2016orbital}%
  \BibitemOpen
  \bibfield  {author} {\bibinfo {author} {\bibfnamefont {L.}~\bibnamefont
  {Zhu}}, \bibinfo {author} {\bibfnamefont {S.}~\bibnamefont {Nie}}, \bibinfo
  {author} {\bibfnamefont {P.}~\bibnamefont {Xiong}}, \bibinfo {author}
  {\bibfnamefont {P.}~\bibnamefont {Schlottmann}},\ and\ \bibinfo {author}
  {\bibfnamefont {J.}~\bibnamefont {Zhao}},\ }\href@noop {} {\bibfield
  {journal} {\bibinfo  {journal} {Nature communications}\ }\textbf {\bibinfo
  {volume} {7}},\ \bibinfo {pages} {10817} (\bibinfo {year}
  {2016}{\natexlab{a}})}\BibitemShut {NoStop}%
\bibitem [{\citenamefont {Zhu}\ \emph {et~al.}(2016{\natexlab{b}})\citenamefont
  {Zhu}, \citenamefont {Woltersdorf},\ and\ \citenamefont
  {Zhao}}]{zhu2016observation}%
  \BibitemOpen
  \bibfield  {author} {\bibinfo {author} {\bibfnamefont {L.}~\bibnamefont
  {Zhu}}, \bibinfo {author} {\bibfnamefont {G.}~\bibnamefont {Woltersdorf}},\
  and\ \bibinfo {author} {\bibfnamefont {J.}~\bibnamefont {Zhao}},\ }\href@noop
  {} {\bibfield  {journal} {\bibinfo  {journal} {Scientific reports}\ }\textbf
  {\bibinfo {volume} {6}},\ \bibinfo {pages} {34549} (\bibinfo {year}
  {2016}{\natexlab{b}})}\BibitemShut {NoStop}%
\bibitem [{\citenamefont {Zawadowski}(1980)}]{zawadowski1980kondo}%
  \BibitemOpen
  \bibfield  {author} {\bibinfo {author} {\bibfnamefont {A.}~\bibnamefont
  {Zawadowski}},\ }\href@noop {} {\bibfield  {journal} {\bibinfo  {journal}
  {Physical Review Letters}\ }\textbf {\bibinfo {volume} {45}},\ \bibinfo
  {pages} {211} (\bibinfo {year} {1980})}\BibitemShut {NoStop}%
\bibitem [{\citenamefont {Cox}\ and\ \citenamefont
  {Zawadowski}(1998)}]{cox1997exotic}%
  \BibitemOpen
  \bibfield  {author} {\bibinfo {author} {\bibfnamefont {D.}~\bibnamefont
  {Cox}}\ and\ \bibinfo {author} {\bibfnamefont {A.}~\bibnamefont
  {Zawadowski}},\ }\href@noop {} {\bibfield  {journal} {\bibinfo  {journal}
  {Adv. Phys.}\ }\textbf {\bibinfo {volume} {47}},\ \bibinfo {pages} {599}
  (\bibinfo {year} {1998})}\BibitemShut {NoStop}%
\bibitem [{\citenamefont {von Delft}\ \emph {et~al.}(1998)\citenamefont {von
  Delft}, \citenamefont {Ralph}, \citenamefont {Buhrman}, \citenamefont
  {Upadhyay}, \citenamefont {Louie}, \citenamefont {Ludwig},\ and\
  \citenamefont {Ambegaokar}}]{von19982}%
  \BibitemOpen
  \bibfield  {author} {\bibinfo {author} {\bibfnamefont {J.}~\bibnamefont {von
  Delft}}, \bibinfo {author} {\bibfnamefont {D.}~\bibnamefont {Ralph}},
  \bibinfo {author} {\bibfnamefont {R.}~\bibnamefont {Buhrman}}, \bibinfo
  {author} {\bibfnamefont {S.}~\bibnamefont {Upadhyay}}, \bibinfo {author}
  {\bibfnamefont {R.}~\bibnamefont {Louie}}, \bibinfo {author} {\bibfnamefont
  {A.}~\bibnamefont {Ludwig}},\ and\ \bibinfo {author} {\bibfnamefont
  {V.}~\bibnamefont {Ambegaokar}},\ }\href@noop {} {\bibfield  {journal}
  {\bibinfo  {journal} {Annals of Physics}\ }\textbf {\bibinfo {volume}
  {263}},\ \bibinfo {pages} {1} (\bibinfo {year} {1998})}\BibitemShut {NoStop}%
\bibitem [{\citenamefont {Abdel-Hafiez}\ \emph {et~al.}(2016)\citenamefont
  {Abdel-Hafiez}, \citenamefont {Zhao}, \citenamefont {Kordyuk}, \citenamefont
  {Fang}, \citenamefont {Pan}, \citenamefont {He}, \citenamefont {Duan},
  \citenamefont {Zhao},\ and\ \citenamefont {Chen}}]{abdel2016enhancement}%
  \BibitemOpen
  \bibfield  {author} {\bibinfo {author} {\bibfnamefont {M.}~\bibnamefont
  {Abdel-Hafiez}}, \bibinfo {author} {\bibfnamefont {X.-M.}\ \bibnamefont
  {Zhao}}, \bibinfo {author} {\bibfnamefont {A.}~\bibnamefont {Kordyuk}},
  \bibinfo {author} {\bibfnamefont {Y.-W.}\ \bibnamefont {Fang}}, \bibinfo
  {author} {\bibfnamefont {B.}~\bibnamefont {Pan}}, \bibinfo {author}
  {\bibfnamefont {Z.}~\bibnamefont {He}}, \bibinfo {author} {\bibfnamefont
  {C.-G.}\ \bibnamefont {Duan}}, \bibinfo {author} {\bibfnamefont
  {J.}~\bibnamefont {Zhao}},\ and\ \bibinfo {author} {\bibfnamefont {X.-J.}\
  \bibnamefont {Chen}},\ }\href@noop {} {\bibfield  {journal} {\bibinfo
  {journal} {Scientific reports}\ }\textbf {\bibinfo {volume} {6}},\ \bibinfo
  {pages} {31824} (\bibinfo {year} {2016})}\BibitemShut {NoStop}%
\bibitem [{\citenamefont {Anderson}\ \emph {et~al.}(1972)\citenamefont
  {Anderson}, \citenamefont {Halperin},\ and\ \citenamefont
  {Varma}}]{anderson1972anomalous}%
  \BibitemOpen
  \bibfield  {author} {\bibinfo {author} {\bibfnamefont {P.~W.}\ \bibnamefont
  {Anderson}}, \bibinfo {author} {\bibfnamefont {B.~I.}\ \bibnamefont
  {Halperin}},\ and\ \bibinfo {author} {\bibfnamefont {C.~M.}\ \bibnamefont
  {Varma}},\ }\href@noop {} {\bibfield  {journal} {\bibinfo  {journal}
  {Philosophical Magazine}\ }\textbf {\bibinfo {volume} {25}},\ \bibinfo
  {pages} {1} (\bibinfo {year} {1972})}\BibitemShut {NoStop}%
\bibitem [{\citenamefont {Phillips}(1972)}]{phillips1972tunneling}%
  \BibitemOpen
  \bibfield  {author} {\bibinfo {author} {\bibfnamefont {W.~A.}\ \bibnamefont
  {Phillips}},\ }\href@noop {} {\bibfield  {journal} {\bibinfo  {journal}
  {Journal of low temperature physics}\ }\textbf {\bibinfo {volume} {7}},\
  \bibinfo {pages} {351} (\bibinfo {year} {1972})}\BibitemShut {NoStop}%
\bibitem [{\citenamefont {Chen}\ \emph {et~al.}(2016)\citenamefont {Chen},
  \citenamefont {Chikara}, \citenamefont {Zapf},\ and\ \citenamefont
  {Morosan}}]{chen2016correlations}%
  \BibitemOpen
  \bibfield  {author} {\bibinfo {author} {\bibfnamefont {C.-W.}\ \bibnamefont
  {Chen}}, \bibinfo {author} {\bibfnamefont {S.}~\bibnamefont {Chikara}},
  \bibinfo {author} {\bibfnamefont {V.~S.}\ \bibnamefont {Zapf}},\ and\
  \bibinfo {author} {\bibfnamefont {E.}~\bibnamefont {Morosan}},\ }\href@noop
  {} {\bibfield  {journal} {\bibinfo  {journal} {Physical Review B}\ }\textbf
  {\bibinfo {volume} {94}},\ \bibinfo {pages} {054406} (\bibinfo {year}
  {2016})}\BibitemShut {NoStop}%
\bibitem [{\citenamefont {Czulucki}\ \emph {et~al.}(2010)\citenamefont
  {Czulucki}, \citenamefont {Auffermann}, \citenamefont {Bednarski},
  \citenamefont {Bochenek}, \citenamefont {B{\"o}hme}, \citenamefont
  {Cichorek}, \citenamefont {Niewa}, \citenamefont {Oeschler}, \citenamefont
  {Schmidt}, \citenamefont {Steglich} \emph {et~al.}}]{czulucki2010crystal}%
  \BibitemOpen
  \bibfield  {author} {\bibinfo {author} {\bibfnamefont {A.}~\bibnamefont
  {Czulucki}}, \bibinfo {author} {\bibfnamefont {G.}~\bibnamefont
  {Auffermann}}, \bibinfo {author} {\bibfnamefont {M.}~\bibnamefont
  {Bednarski}}, \bibinfo {author} {\bibfnamefont {{\L}.}~\bibnamefont
  {Bochenek}}, \bibinfo {author} {\bibfnamefont {M.}~\bibnamefont {B{\"o}hme}},
  \bibinfo {author} {\bibfnamefont {T.}~\bibnamefont {Cichorek}}, \bibinfo
  {author} {\bibfnamefont {R.}~\bibnamefont {Niewa}}, \bibinfo {author}
  {\bibfnamefont {N.}~\bibnamefont {Oeschler}}, \bibinfo {author}
  {\bibfnamefont {M.}~\bibnamefont {Schmidt}}, \bibinfo {author} {\bibfnamefont
  {F.}~\bibnamefont {Steglich}}, \emph {et~al.},\ }\href@noop {} {\bibfield
  {journal} {\bibinfo  {journal} {ChemPhysChem}\ }\textbf {\bibinfo {volume}
  {11}},\ \bibinfo {pages} {2639} (\bibinfo {year} {2010})}\BibitemShut
  {NoStop}%
\bibitem [{\citenamefont {Vladar}\ and\ \citenamefont
  {Zawadowski}(1983)}]{vladar1983theory1}%
  \BibitemOpen
  \bibfield  {author} {\bibinfo {author} {\bibfnamefont {K.}~\bibnamefont
  {Vladar}}\ and\ \bibinfo {author} {\bibfnamefont {A.}~\bibnamefont
  {Zawadowski}},\ }\href@noop {} {\bibfield  {journal} {\bibinfo  {journal}
  {Physical Review B}\ }\textbf {\bibinfo {volume} {28}},\ \bibinfo {pages}
  {1564} (\bibinfo {year} {1983})}\BibitemShut {NoStop}%
\bibitem [{\citenamefont {Vlad{\'a}r}\ and\ \citenamefont
  {Zawadowski}(1983)}]{vladar1983theory2}%
  \BibitemOpen
  \bibfield  {author} {\bibinfo {author} {\bibfnamefont {K.}~\bibnamefont
  {Vlad{\'a}r}}\ and\ \bibinfo {author} {\bibfnamefont {A.}~\bibnamefont
  {Zawadowski}},\ }\href@noop {} {\bibfield  {journal} {\bibinfo  {journal}
  {Physical Review B}\ }\textbf {\bibinfo {volume} {28}},\ \bibinfo {pages}
  {1596} (\bibinfo {year} {1983})}\BibitemShut {NoStop}%
\bibitem [{\citenamefont {Cichorek}\ \emph {et~al.}(2017)\citenamefont
  {Cichorek}, \citenamefont {Bochenek}, \citenamefont {Schmidt}, \citenamefont
  {Niewa}, \citenamefont {Czulucki}, \citenamefont {Auffermann}, \citenamefont
  {Steglich}, \citenamefont {Kniep},\ and\ \citenamefont
  {Kirchner}}]{cichorek2017cichorek}%
  \BibitemOpen
  \bibfield  {author} {\bibinfo {author} {\bibfnamefont {T.}~\bibnamefont
  {Cichorek}}, \bibinfo {author} {\bibfnamefont {L.}~\bibnamefont {Bochenek}},
  \bibinfo {author} {\bibfnamefont {M.}~\bibnamefont {Schmidt}}, \bibinfo
  {author} {\bibfnamefont {R.}~\bibnamefont {Niewa}}, \bibinfo {author}
  {\bibfnamefont {A.}~\bibnamefont {Czulucki}}, \bibinfo {author}
  {\bibfnamefont {G.}~\bibnamefont {Auffermann}}, \bibinfo {author}
  {\bibfnamefont {F.}~\bibnamefont {Steglich}}, \bibinfo {author}
  {\bibfnamefont {R.}~\bibnamefont {Kniep}},\ and\ \bibinfo {author}
  {\bibfnamefont {S.}~\bibnamefont {Kirchner}},\ }\href@noop {} {\bibfield
  {journal} {\bibinfo  {journal} {Physical Review Letters}\ }\textbf {\bibinfo
  {volume} {118}},\ \bibinfo {pages} {259702} (\bibinfo {year}
  {2017})}\BibitemShut {NoStop}%
\end{thebibliography}%
\bibliographystyle{apsrev4-2}

\end{document}